\documentclass[traditabstract, printer]{aa}

\usepackage{color}
\usepackage{xcolor}
\usepackage{pifont}
\usepackage{txfonts}
\usepackage{graphicx}
\usepackage{hyperref}
\usepackage{soul}
\hypersetup{
    pdftoolbar=true,			
    pdfmenubar=true,			
    pdffitwindow=false,			
    pdfstartview={FitH},		
    pdftitle={title},			
    pdfauthor={Author},			
    pdfsubject={Subject},		
    pdfcreator={Author},		
    pdfproducer={Author},		
    pdfkeywords={Population \& Evolutionary} ,        
    pdfnewwindow=true,			
    colorlinks=true,			
    linkcolor=cyan,				
    citecolor=blue,				
    filecolor=cyan,				
    urlcolor=cyan				
}

\newcommand{\Fado}{{\sc Fado}}
\newcommand{\SL}{{\sc Starlight}}
\newcommand{\Steckmap}{{\sc Steckmap}}

\def\EWha{EW(H$\alpha$)}

\newcommand{\FadoFCmode}{{\sc Fado}-$\mathtt{FC}$mode}
\newcommand{\FadoSTmode}{{\sc Fado}-$\mathtt{ST}$mode}
\begin{document}

\title{Revisiting stellar properties of star-forming galaxies with stellar and nebular spectral modelling}

\author{
Leandro S. M. Cardoso \inst{\ref{adress1}}
\and Jean Michel Gomes \inst{\ref{adress1}}
\and Polychronis Papaderos\inst{\ref{adress1},\ref{adress2},\ref{adress3}}
\and Ciro Pappalardo\inst{\ref{adress2},\ref{adress3}}
\and Henrique Miranda\inst{\ref{adress2},\ref{adress3}}
\and Ana Paulino-Afonso\inst{\ref{adress1}}
\and José Afonso\inst{\ref{adress2},\ref{adress3}}
\and Patricio Lagos\inst{\ref{adress1}}
}

\institute{
Instituto de Astrof\' isica e Ci\^ encias do Espa\c co, Universidade do Porto, CAUP, Rua das Estrelas, PT4150-762 Porto, Portugal\label{adress1} 
\and Instituto de Astrof\' isica e Ci\^ encias do Espa\c co, Universidade de Lisboa, OAL, Tapada da Ajuda, PT1349-018 Lisboa, Portugal\label{adress2} 
\and Departamento de Física, Faculdade de Ciências da Universidade de Lisboa, Edifício C8, Campo Grande, PT1749-016 Lisboa, Portugal\label{adress3}
}
\date{Received ?? / Accepted ??}

\abstract
{Spectral synthesis is a powerful tool for interpreting the physical properties of galaxies by decomposing their spectral energy distributions (SEDs) into the main luminosity contributors (e.g. stellar populations of distinct age and metallicity or ionised gas). However, the impact nebular emission has on the inferred properties of star-forming (SF) galaxies has been largely overlooked over the years, with unknown ramifications to the current understanding of galaxy  evolution.}
{ The objective of this work is to estimate the relations between stellar properties (e.g. total mass, mean age, and mean metallicity) of SF galaxies by simultaneously fitting the stellar and nebular continua and comparing them to the results derived through the more common purely stellar spectral synthesis approach. }
{ The main galaxy sample from SDSS DR7 was analysed with two distinct population synthesis codes: \Fado, which estimates self-consistently both the stellar and nebular contributions to the SED, and the original version of \SL, as representative of purely stellar population synthesis codes.  }
{ Differences between codes regarding average mass, mean age and mean metallicity values can go as high as $\sim$0.06 dex for the overall population of galaxies and $\sim$0.12 dex for SF galaxies (galaxies with EW(H$\alpha$)>3 \AA), with the most prominent difference between both codes in the two populations being in the light-weighted mean stellar age. \Fado\ presents a broader range of mean stellar ages and metallicities for SF galaxies than \SL, with the latter code preferring metallicity solutions around the solar value ($Z_{\odot} = 0.02$).  A closer look into the average light- and mass-weighted star formation histories of intensively SF galaxies (EW(H$\alpha$)>75 \AA) reveals that the light contributions of simple stellar populations (SSPs) younger than $\leq 10^7$ ($10^9$) years in \SL\ are higher by $\sim$5.41\% (9.11\%) compared to \Fado. Moreover, \Fado\ presents higher light contributions from SSPs with metallicity $\leq Z_\odot / 200$ ($Z_\odot / 50$) of around 8.05\% (13.51\%) when compared with \SL. This suggests that \SL\ is underestimating the average light-weighted age of intensively SF galaxies by up to $\sim$0.17 dex and overestimating the light-weighted metallicity by up to $\sim$0.13 dex compared to \Fado\ (or vice versa). The comparison between the average stellar properties of passive, SF and intensively SF galaxy samples also reveals that differences between codes increase with increasing EW(H$\alpha$) and decreasing total stellar mass. Moreover, comparing SF results from \Fado\ in a purely stellar mode with the previous results qualitatively suggests that differences between codes are primarily due to mathematical and statistical differences and secondarily due to the impact of the nebular continuum modelling approach (or lack thereof). However, it is challenging to adequately quantify the relative role of each factor since they are likely interconnected. }
{ This work finds indirect evidence that a purely stellar population synthesis approach negatively impacts the inferred stellar properties (e.g. mean age and mean metallicity) of galaxies with relatively high star formation rates (e.g. dwarf spirals, `green peas', and starburst galaxies). In turn, this can bias interpretations of fundamental relations such as the mass-age or mass-metallicity, which are factors worth bearing in mind in light of future high-resolution  spectroscopic surveys at higher redshifts (e.g. MOONS and 4MOST-4HS). }

\maketitle

\section{Introduction}\label{Section_-_Introduction}

	Understanding the complex physical processes behind Galaxy formation and evolution can be a daunting endeavour, especially with the increasing technical difficulties when looking further back in time. However, several keystone results have been obtained over the past decades: (a) high-mass galaxies assemble most of their stellar content early in their lives, whereas low-mass galaxies display relatively high specific star-formation rates (sSFRs) at a late cosmic epoch (e.g. \citealt{Brinchmann_etal_2004, Noeske_etal_2007a}), a phenomenon known as galaxy downsizing (e.g. \citealt{Cowie_etal_1996, Heavens_etal_2004, Cimatti_Daddi_Renzini_2006}), (b) there is a relative tight correlation between the stellar mass and gas-phase metallicity (e.g. \citealt{Lequeux_etal_1979, Tremonti_etal_2004}), (c) galaxies display a bimodal distribution based on their colour, in which they can either be large, concentrated, red and quiescent \emph{or} small, less concentrated, blue and presenting signs of recent star formation (e.g. \citealt{Gladders_Yee_2000, Strateva_etal_2001, Blanton_etal_2003, Kauffmann_etal_2003b, Baldry_etal_2004, Mateus_etal_2006}), and (d) the seemingly linear correlation between the stellar mass and SFR in star-forming (SF) galaxies, an observable that is usually referred to as the SF main sequence of galaxies (e.g. \citealt{Brinchmann_etal_2004, Noeske_etal_2007a}). 

	Notwithstanding these insights, the overall picture of galaxy formation and evolution is far from complete. For instance, a persistent issue in interpreting the physical characteristics of SF galaxies from spectral synthesis has been the relative contribution of nebular emission and its potential impact on the estimated stellar and overall galaxy properties. Indeed, several studies have shown that nebular emission (continuum and lines) can account up to $\sim$60\% of the monochromatic luminosity at $\sim$5000 \AA\ in AGN and starburst galaxies (e.g. \citealt{Krueger_etal_1995, Papaderos_etal_1998, Zackrisson_etal_2001, Zackrisson_Bergvall_Leitet_2008, Schaerer_deBarros_2009,  Papaderos_Ostlin_2012}) and in optical broadband photometry, specially at low metallicities (\citealt{Anders_FritzeAlvensleben_2003}). 

	For instance, \cite{Krueger_etal_1995} noted that strong star-formation activity can lead nebular continuum to contribute $\sim$30--70\% to the total optical and near-infrared emission, whereas emission lines alone account for up to $\sim$45\% in optical broadband photometry. From a different perspective, \cite{Reines_etal_2010} carried out spectral modelling of two young massive star clusters in NGC 4449 and showed that both the nebular continuum and line emission have a major impact on the estimated magnitudes and colours of young clusters ($\lesssim$ 5 Myr), thus also affecting their age, mass and extinction estimates. Moreover, \cite{Izotov_Guseva_Thuan_2011} studied a sample of 803 SF luminous compact galaxies (i.e. `green peas') from the SDSS (\citealt{York_etal_2000}) and warned that a purely stellar spectral modelling of such objects can lead to the overestimation of the relative contribution of old stellar populations by as much as a factor of four. As the authors noted, there are two reasons for this overestimation: (a) nebular continuum increases the galaxy luminosity and (b) the nebular continuum is flatter than the stellar, thus making the overall continuum redder that it would be if it were purely stellar. \cite{Papaderos_Ostlin_2012} also showed that nebular emission can introduce strong photometric biases of 0.4-1 mag in galaxies with high specific SFRs, whereas  \cite{Pacifici_etal_2015} found that that SFRs can be overestimated by up $\sim$0.12 dex if nebular emission is neglected in spectral modelling. 

	The impact of these biases on the stellar mass and gas-phase metallicity relation (e.g. \citealt{Tremonti_etal_2004}), SF main sequence (e.g. \citealt{Brinchmann_etal_2004, Noeske_etal_2007a}) or other scaling relations involving stellar properties is still largely unexplored. In fact, it is important to note that most spectral modelling of large-scale surveys using population synthesis has been carried assuming a purely stellar modelling approach, regardless of wether it is applied to passive, SF or even active galaxies  (e.g. \citealt{Kauffmann_etal_2003a, Panter_Heavens_Jimenez_2003, CidFernandes_etal_2005, Asari_etal_2007, Tojeiro_etal_2009, Zhong_etal_2010,  TorresPapaqui_etal_2012, Perez_etal_2013, SanchezBlazquez_etal_2014, LopezFernandez_etal_2016, Rosa_etal_2018, Kuzmicz_Czerny_Wildy_2019, Cai_etal_2020, Cai_Zhao_Bai_2021}).

	Although nebular continuum and line emission has been adopted in several evolutionary synthesis models (e.g. \citealt{Leitherer_etal_1999, Schaerer_2002, Molla_GarciaVargas_Bressan_2009, MartinManjon_etal_2010}), no consistent nebular prescription has been implemented in inversion population synthesis codes until recently. Indeed, \cite{Gomes_Papaderos_2017, Gomes_Papaderos_2018} presented in \Fado\ the first population synthesis code to fit self-consistently both the stellar and nebular spectral components while adopting a genetic optimisation framework. Tests revealed that this code estimates the main stellar population properties of SF galaxies (e.g. mass, mean age, and mean metallicity) within an accuracy of $\sim$0.2 dex (\citealt{Gomes_Papaderos_2017, Gomes_Papaderos_2018, Cardoso_Gomes_Papaderos_2019, Pappalardo_etal_2021}). For comparison, the typical level of deviations between input and inferred stellar properties when applying purely stellar population synthesis codes to evolved stellar populations with faint or absent nebular emission is of $\sim$0.15-0.2 dex (e.g. \citealt{CidFernandes_etal_2005, CidFernandes_etal_2014, Ocvirk_etal_2006a, Ocvirk_etal_2006b, Tojeiro_etal_2007, Tojeiro_etal_2009, Koleva_etal_2009}). 
	
	Moreover, \cite{Cardoso_Gomes_Papaderos_2019} (hereafter CGP19) compared \Fado\ with a purely stellar population synthesis code using synthetic galaxy models for different star formation histories (SFHs; e.g. instantaneous burst, continuous, and exponentially declining) and different fitting configurations (e.g. with or without emission lines and including or excluding the Balmer and Paschen continuum discontinuities at 3646 and 8207 \AA, respectively). This work showed that applying the public version of the purely stellar population synthesis code \SL\footnote{The same version adopted in this work. Not to be confused with the one introduced by \cite{LopezFernandez_etal_2016}, which combines UV photometry with spectral fitting in the optical.} (\citealt{CidFernandes_etal_2005}) to spectra with a relatively high nebular continuum  can lead to the overestimation of the total stellar mass by as much as $\sim$2 dex and the mass-weighted mean stellar age up to $\sim$4 dex, whereas the mean metallicity and light-weighted mean stellar age can both be underestimated by up to $\sim$0.6 dex. Moreover, it was found that these stellar properties can still be recovered with \Fado\ within $\sim$0.2 dex in evolutionary stages with severe nebular contamination. \cite{Pappalardo_etal_2021} (hereafter P21) continued this line of inquiry by adding the non-parametric purely stellar \Steckmap\ (\citealt{Ocvirk_etal_2006a, Ocvirk_etal_2006b}) to the code comparison while also exploring the impact of varying spectral quality on the derived physical properties, finding similar results and trends. It is important to note that (a) these tests were carried out using the same evolutionary models (\citealt{Bruzual_Charlot_2003}) both in the creation of the synthetic spectra and in the spectral modelling and (b) the input synthetic composite stellar populations were built having a constant solar metallicity ($Z_\odot=0.02$). Thus, these uncertainties should be viewed as upper limits and might not necessarily directly translate into biases affecting observations. 

	Also recently, \cite{Gunawardhana_etal_2020} studied the stellar and nebular characteristics of massive stellar populations in the Antennae galaxy using an updated version of \textsc{Platefit} (\citealt{Tremonti_etal_2004, Brinchmann_etal_2004}), capable of self-consistent modelling of the  stellar and nebular continua. Spectral fitting provides estimates of the stellar and gas metallicities, stellar ages and electron temperature $T_e$ and density $n_e$ by taking as reference model libraries of HII regions built using the evolutionary synthesis code \textsc{Starburst99} (\citealt{Leitherer_etal_1999}) and the photoionisation code \textsc{Cloudy} (\citealt{Ferland_etal_1998,Ferland_etal_2013}). This work found the stellar and gas metallicity of the starbursts to be near solar and that the metallicity of the star-forming gas in the loop of NGC 4038 appears to be slightly richer than the rest of the galaxy.

	Using a different approach, \cite{LopezFernandez_etal_2016} presented a new version of the population synthesis code \SL\ (\citealt{CidFernandes_etal_2005}) that combines optical spectroscopy with UV photometry. This work used a mixture of simulated and real CALIFA data (\citealt{Sanchez_etal_2012}) and found that the additional UV constraints have a low impact on the inferred stellar mass and dust optical depth. Although the mean age and metallicity of most galaxies remains unaffected by the additional UV spectral information,  this work also showed that stellar populations of low-mass late-type galaxies are older and less chemically enriched than in purely-optical modelling. \cite{Werle_etal_2019} pursued further this line of inquiry by combining GALEX photometry with SDSS spectroscopy and found that the UV constraints lead to the increase of simple stellar populations (SSPs) with ages between $\sim$10$^7$ and 10$^8$ years in detriment to the relative contribution of younger and older populations, leading to slightly older mean stellar ages when weighted my mass. This redistribution of the SFH is particularly noticeable in galaxies in the low-mass end of the blue cloud. Later, \cite{Werle_etal_2020} adopted a similar approach to study early-type galaxies in the same sample and found that the UV constraints broadens the attenuation, mean stellar age and metallicity distributions. Moreover, galaxies with young stellar populations have larger H$\alpha$ equivalent widths (EWs) and larger attenuations, with the metallicity of these populations being increasingly lower for larger stellar masses. Although these three works successfully use UV spectral information to constrain the contribution of young stellar populations, one wonders if the nebular continuum in galaxies with relatively high specific SFRs still needs to be accounted for as intermediate-to-old stellar populations in purely stellar spectral synthesis, in which case the lack of nebular continuum treatment would still affect the inferred stellar properties.
	
	Taking all these factors into consideration, it is possible that the current understanding of galaxy evolution, specifically that of SF galaxies in the local universe based on large-scale survey analysis, has been affected by a lack of an adequate nebular modelling prescription in previous spectral synthesis codes. To address this subject, this work aims to revisit the relation between key stellar properties (e.g. mass, mean age, and mean metallicity) of SF galaxies by comparing the results obtained with \Fado\ and a representative of purely stellar population synthesis codes (i.e. \SL) when applied to a well-studied large-scale survey such as SDSS (\citealt{York_etal_2000}). 

	This paper is organised as follows. Section \ref{Section_-_Methodology} details the methodology adopted to extract and analyse the SDSS DR7 spectra using spectral synthesis. The main results of this work are presented in Section \ref{Section_-_Results}, which is particularly focussed on the relation between the main stellar properties of SF galaxies when adopting two different spectral modelling approaches. Finally, in Sections \ref{Section_-_Discussion} and \ref{Section_-_Conclusions} the findings of this work are discussed and summarised, respectively. Unless stated otherwise, this work assumes $H_0 = 70$ km s$^{-1}$ Mpc$^{-1}$, $\Omega_M=0.3$ and $\Omega_\Lambda=0.7$.

\section{Methodology}\label{Section_-_Methodology}

	The population synthesis codes \Fado\ and \SL\ were applied to the galaxy sample from SDSS Data Release 7 (\citealt{Abazajian_etal_2009}). This data set contains multi-band spectrophotometric data for 926246 objects, each with a single-fibre integrated spectrum with the wavelength range of 3800-9200 \AA\ at a resolution of $R\sim 1800$--2200, with the fibre covering $\sim$5.5 kpc of the central region of an object at $z \! \sim \! 0.1$. There are several reasons to utilise this dataset: (a) photometric completeness, (b) relatively wide redshift coverage ($0.02 \! \lesssim \! z \! \lesssim \! 0.6$), (c) uniform spectral calibration, and (d) wide-range of topics related to galaxy formation and evolution that have been tackled with this data (e.g. mass-metallicity relation, \citealt{Tremonti_etal_2004}; stellar mass assembly, \citealt{Kauffmann_etal_2003a}; galaxy bimodality, \citealt{Strateva_etal_2001, Blanton_etal_2003, Kauffmann_etal_2003b, Baldry_etal_2004}; host-AGN relation, \citealt{Kauffmann_etal_2003c, Hao_etal_2005}; environmental effects on the colour-magnitude relation \citealt{Hogg_etal_2004, Cooper_etal_2010}).

	The raw spectra were corrected for extinction using the colour excess correction factor $\mathrm{E}(B-V)$ computed based on the \cite{Schlegel_Finkbeiner_Davis_1998} dust maps, considering the objects sky position using the right ascension and declination provided by the survey, combined with the correcting factor to $\mathrm{E}(B-V)$ of \cite{Schlafly_Finkbeiner_2011}. The full correction equation reads,
	
	\begin{equation}
	F_{\lambda}^{corrected} = F_{\lambda}^{raw} \cdot 10^{0.4 \, \cdot \, \mathrm{E}(B-V) \cdot \left( \frac{A_\lambda}{A_V} \right) \cdot R_V}, 
																			\end{equation}			

\noindent where $F_{\lambda}^{corrected}$ and $F_{\lambda}^{raw}$ are the corrected and raw fluxes as a function of wavelength $\lambda$, respectively, and $A_V$ is the extinction in the \emph{V}-band. The adopted extinction curve (also known as the reddening law) was \cite{Cardelli_Clayton_Mathis_1989}, with $R_V = 3.1$. The spectra were then converted to the observer restframe and rebinned so that $\Delta\lambda = 1$ \AA.  
	
	As detailed in \cite{Gomes_Papaderos_2017}, the main task of the population synthesis code \Fado\ is to reproduce the observed spectral energy distribution (SED) through a linear combination of spectral components (e.g. individual stellar spectra or SSPs) as expressed by:

\begin{multline}\label{Equation_-_FADO}
F_\lambda = \sum_{i=1}^{N_\star} M_{i,\lambda_0} \cdot L_{i,\lambda} \cdot 10^{-0.4 \cdot A_V \cdot q_\lambda} \otimes S( v_\star, \sigma_\star )  \\
          + \Gamma_\lambda(n_e,T_e) \cdot 10^{-0.4 \cdot A_V^{neb} \cdot q_\lambda} \otimes N(v_\eta,\sigma_\eta) ,
																	\end{multline}
																	
\noindent where $F_\lambda$ is the flux of the observed spectrum, $N_\star$ is the number of unique spectral components in the adopted base library, $M_{i,\lambda_0}$ is the stellar mass of the $i^{\mathrm{th}}$ spectral component at the normalisation wavelength $\lambda_0$, $L_{j,\lambda}$ is the luminosity contribution of the $i^{\mathrm{th}}$ spectral component, $A_V$ is the \emph{V}-band extinction, $q_\lambda$ is the ratio of $A_\lambda$ over $A_V$, $S( v_\star, \sigma_\star )$ denotes a Guassian kernel simulating the effect of stellar kinematics on the spectrum, with $v_\star$ and $\sigma_\star$ representing the stellar shift and dispersion velocities, respectively, $\Gamma_\lambda(n_e,T_e)$ is the nebular continuum computed assuming that all stellar photons with $\lambda \leq 911.76$ \AA\ are absorbed and reprocessed into nebular emission, under the supposition that case B recombination applies, $A_V^{neb}$ is the nebular \emph{V}-band extinction, and $N(v_\eta,\sigma_\eta)$ denotes the nebular kinematics kernel, with $v_\eta$ and $\sigma_\eta$ representing the nebular shift and dispersion velocities, respectively. 

	It is important to note that all publicly available population synthesis codes before \Fado\ aimed to reconstruct the observed continuum using only the first term on the right-hand side of Equation \ref{Equation_-_FADO}, meaning, using a purely stellar scheme to reconstruct the overall observed SED (more details in \citealt{Gomes_Papaderos_2017, Gomes_Papaderos_2018}). Indeed, one of the innovative features of \Fado\ is the inclusion of the second term, which represents the relative spectral contribution of the nebular continuum inferred from the Lyman continuum production rate of the young stellar component and scaling with the intensity of prominent Balmer lines. This makes \Fado\ a tool specially designed to self-consistently model the stellar and nebular continuum components of SF and starburst galaxies. 
      
	Subsequently spectral fitting was carried out with both \Fado\footnote{Version 1b: \url{http://www.spectralsynthesis.org}} and \SL\footnote{Version 4: \url{http://www.starlight.ufsc.br}} between  3400 and 8900 \AA\ using a base library with 150 SSPs from \cite{Bruzual_Charlot_2003} with a \cite{Chabrier_2003} IMF and Padova 1994 evolutionary tracks (\citealt{Alongi_etal_1993, Bressan_etal_1993, Fagotto_etal_1994a, Fagotto_etal_1994b, Girardi_etal_1996}). This stellar library contains 25 ages (between 1 Myr and 15 Gyr) and six metallicities ($Z = 0.0001; 0.0004; 0.004; 0.008; 0.02; 0.05 =  Z_\odot \times \{1/200; 1/50; 1/5; 2/5; 1; 2.5\}$) and corresponds to an expanded version of the base adopted in CGP19, with the addition of $Z_\odot / 200$ and $Z_\odot / 50$ at the lower end of the metallicity range. The spectra were modelled assuming a \cite{Calzetti_etal_2000} extinction law, which was originally constructed taking into consideration integrated observations of SF galaxies. Moreover, the \emph{V}-band extinction, stellar velocity shift and dispersion free parameters where allowed to vary in both codes within the ranges of $A_V = 0$--4 mag, $v_\star =$ -500--500 km/s and $\sigma_\star = 0$--500 km/s, respectively. Identical parameter ranges for the velocity shift and dispersion were adopted for the nebular component in the \Fado\ runs. Moreover, two complete runs of the whole sample were performed with \SL\ while changing the initial guesses in the parameter space in order to evaluate the formal errors (e.g. \citealt{Ribeiro_etal_2016, Cardoso_Gomes_Papaderos_2016, Cardoso_Gomes_Papaderos_2017, Cardoso_Gomes_Papaderos_2019}).
	
	Moreover, a pre-fitting analysis was carried out with \SL\ using a smaller base library with 45 SSPs for 15 ages (between 1 Myr and 13 Gyr) and three metallicities ($Z = \{ 0.004; 0.02; 0.05 \} =  Z_\odot \times \{ 1/5; 1; 2.5 \}$) that comes with the distribution package of \SL\ (c.f. \citealt{CidFernandes_etal_2005}). The results of this preliminary step were used to create individual spectral masks in order to  exclude the most prominent emission lines in the main fitting. This was achieved by subtracting the continuum of the best-fit model from the observation and, from the resulting residual spectrum, by fitting Gaussians to the lines. The masking of these spectral regions ensures that the \SL\ results  are more robust when it comes to SF galaxies, a procedure similar to that adopted by \cite{Asari_etal_2007} and \cite{Ribeiro_etal_2016}.  At the same time, \Fado\ performs the masking of the most prominent emission lines as part of one of its built-in pre-fitting routines, as detailed in  \cite{Gomes_Papaderos_2017}. Three additional spectral regions are masked in both codes which are associated with small bugs in the adopted evolutionary models: 6845--6945, 7165--7210 and 7550--7725 \AA\ (cf. \citealt{Bruzual_Charlot_2003} for more details).

	In addition to masking, spectral modelling was carried out while taking into consideration several spectral flags provided by the SDSS survey that are included in the mask array of each `\emph{fit}' file. These flags mark spectral regions with a wide range of {potential} problems (e.g. no observation, poor calibration, bad pixel, or sky lines) and, therefore, must be excluded from spectral synthesis. The following individual flags provided by the survey were adopted in this work: `Bad pixel within 3 pixels of trace', `Pixel fully rejected in extraction', `Sky level > flux + 10*(flux error)', and `Emission line detected here' (with the later being adopted only for \SL, for reasons previously mentioned). Choosing which individual flags to include when every pixel is important in a pixel-by-pixel code becomes an exercise in compromise. On the one hand, masking spectral regions affected by bad pixels or sky noise leads to more accurate estimates of the physical properties inferred through spectral synthesis. On the other hand, using all spectral flags computed by an automatic pipeline in any large-scale survey can lead to severe over-flagging and,  thus, the removal of important spectral features. The chosen flags reflect this balancing act. 

\begin{figure*}[!t] 
\begin{center}	
\includegraphics[width=\textwidth]{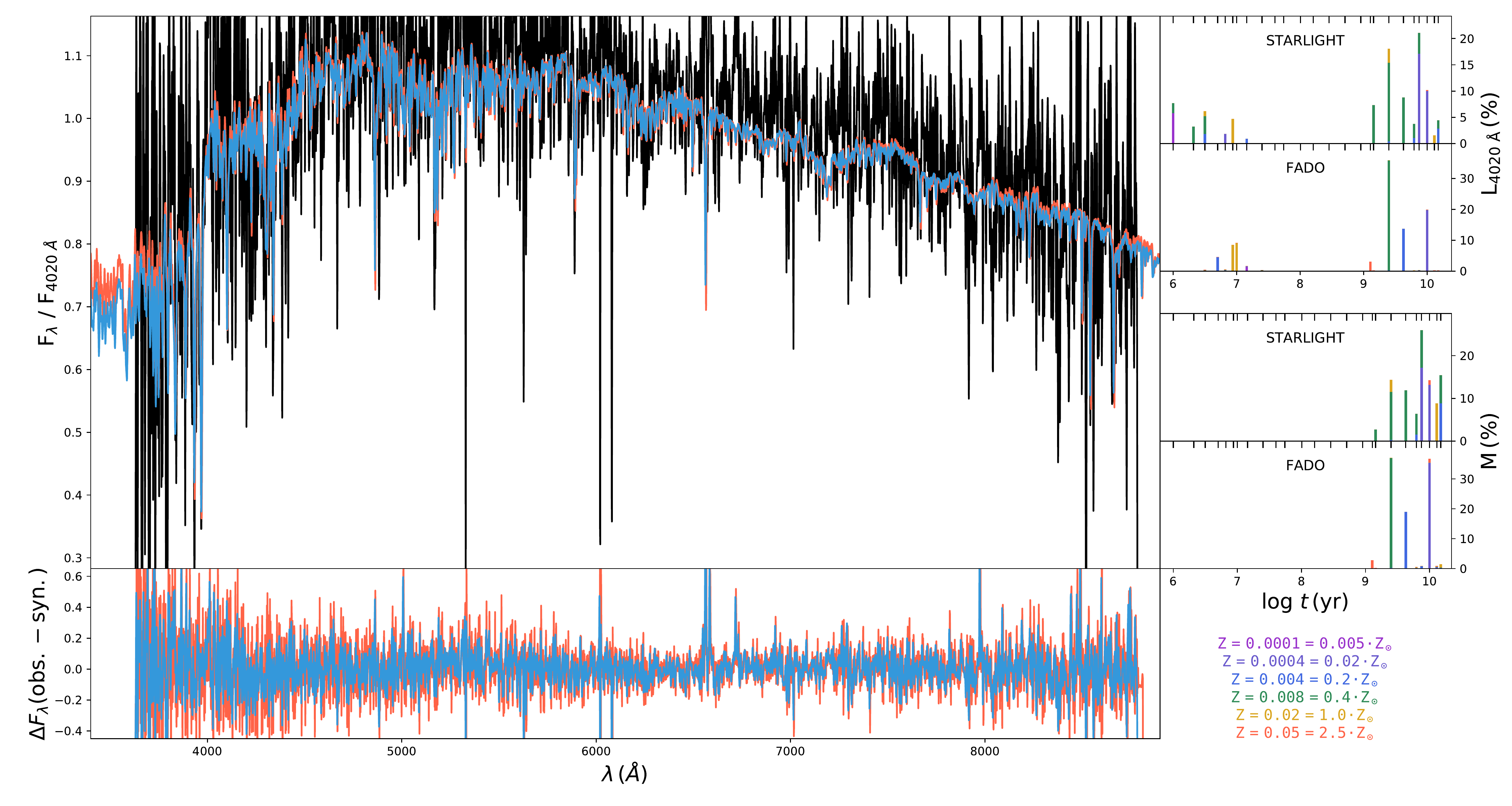}
\caption{Spectral modelling results for the SDSS DR 7 spectrum `52912-1424-151' using the spectral synthesis codes \Fado\ and \SL\ . Main panel: Black, blue and red lines represent the galaxy input spectrum and the best-fit solutions from \Fado\ and \SL, respectively. Bottom panel: Blue and red lines represent the residual spectra of \Fado\ and \SL, respectively, that result from subtracting the best-fit solutions from the input spectrum. Top right-hand side panels: Light fractions $L_{4020 \AA}$ at the normalisation wavelength $\lambda_0 = 4020$ \AA\ as a function of the age of the SSPs in the best-fit solution of \SL\ (top) and \Fado\ (bottom), with colour coding representing metallicity. Bottom right-hand side panels: Mass fractions of the SSPs fitted by \SL\ (top) and \Fado\ (bottom). The black vertical ticks on the top of the x-axes represent the age coverage of the adopted stellar library.  }
\label{Fig_-_text_fit_example}
\end{center}
\end{figure*}

	Figure \ref{Fig_-_text_fit_example} shows an example of spectral fits from both codes for the object `52912-1424-151', a particularly noisy spectrum with $\mathrm{S/N}({\lambda_0}) = 4.4$. Black, red, and blue lines on the main panel represent the input, \SL\ and \Fado\ best-fit spectra, respectively. The right-hand side panels illustrate the best-fit SFHs based on the luminosity contribution $L_{4020}$ of the selected SSPs at the normalisation wavelength $\lambda_0 = 4020$ \AA\ (top panels) and on their corresponding mass contributions (bottom panels). Although the mass distributions obtained with both codes differ very little, with most mass coming from SSPs older than 1 Gyr, \SL\ best-fit solution includes more SSPs younger than 10 Myr compared to \Fado\ when it comes to the light distribution. Given the relatively noisy nature of the observation, this type of variations are not necessarily indicative of a strong divergence between the codes. More reliable trends are found when grouping galaxies in populations with similar spectral characteristics, as explored in Sections \ref{Section_-_Results} and \ref{Section_-_Discussion}.

\section{Results}\label{Section_-_Results}

	The following analysis is a comparison between the results from \Fado\ and \SL, particularly regarding the stellar properties of SF galaxies (e.g. mass, mean age, and mean metallicity). Although these results could be compared to the vast number of previous works that analysed the SDSS (e.g. \citealt{Brinchmann_etal_2004, CidFernandes_etal_2005, Tojeiro_etal_2009}), this endeavour is outside of the scope of this work for several reasons. 
	
	For one, this work is a continuation of CGP19 and P21, in that it follows similar methodologies specifically interested in looking into how (not) modelling the nebular continuum in SF galaxies can impact their inferred physical properties. Moreover, it is notably difficult to ascertain the specific details regarding how previous works dealt with spectral extraction and modelling (e.g. dust treatment, spectral masking, and evolutionary ingredients). This obstacle makes direct comparisons difficult and potentially misleading, since no clear route is available to quantify the assumptions behind the different methodologies. 
	
	Last but not least, there is also the issue of aperture effects and corrections. Several works have issued cautionary remarks when attempting to apply photometric-based aperture corrections to physical properties inferred through spectral synthesis (e.g. \citealt{Richards_etal_2016, Gomes_etal_2016a, Gomes_etal_2016b, Green_etal_2017}), whilst others claim that aperture-free properties such as SFR can still be inferred (e.g. \citealt{DuartePuertas_etal_2017}). Since the main objective is to compare results between two different modelling approaches, no aperture corrections are necessary. Therefore, the following discussion focus only on the galaxy surface area covered by the spectroscopic fibre. 
		
\subsection{Sample selection}\label{SubSection_-_Sample_Selection}

\begin{figure*}[!t] 
\begin{center}	
\includegraphics[width=\textwidth]{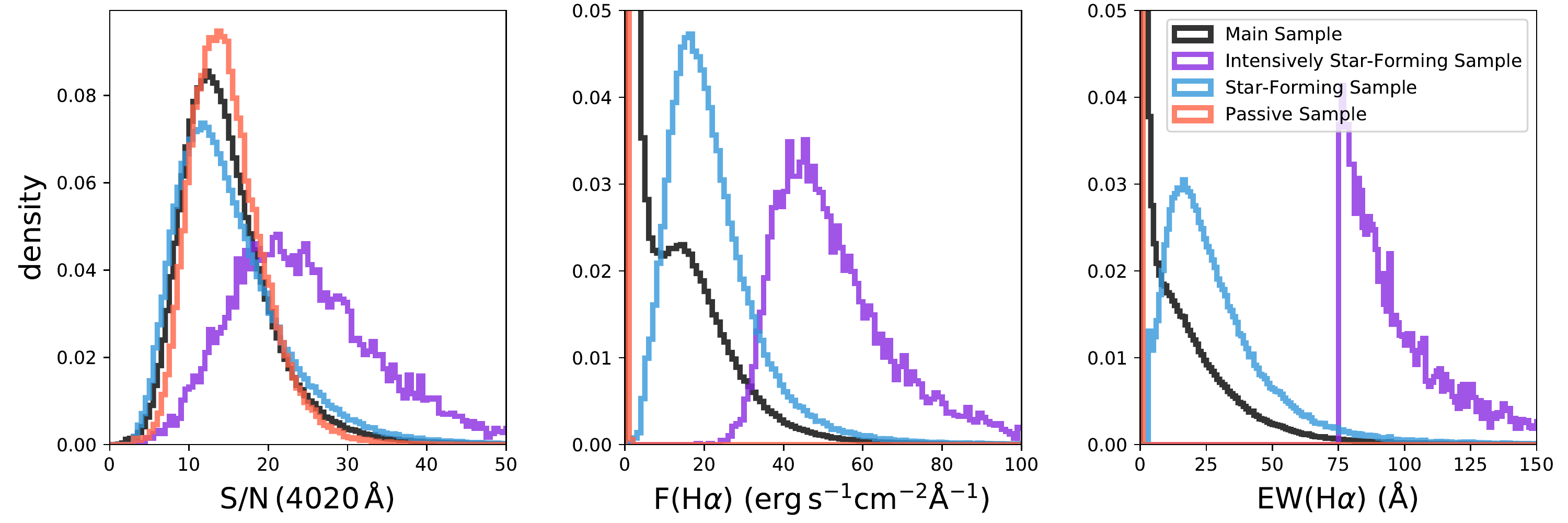}
\caption{Density distributions of the signal-to-noise at the normalisation wavelength S/N(4020\AA), flux of the H$\alpha$ emission line F(H$\alpha$) and H$\alpha$ equivalent width EW(H$\alpha$) of the Main, Intensively Star-forming, Star-forming and Passive Samples represented by the black, violet, blue and red lines, respectively. }
\label{Fig_-_SN_F_EW}
\end{center}
\end{figure*}
\begin{figure} 
\begin{center}	
\includegraphics[width=0.45\textwidth]{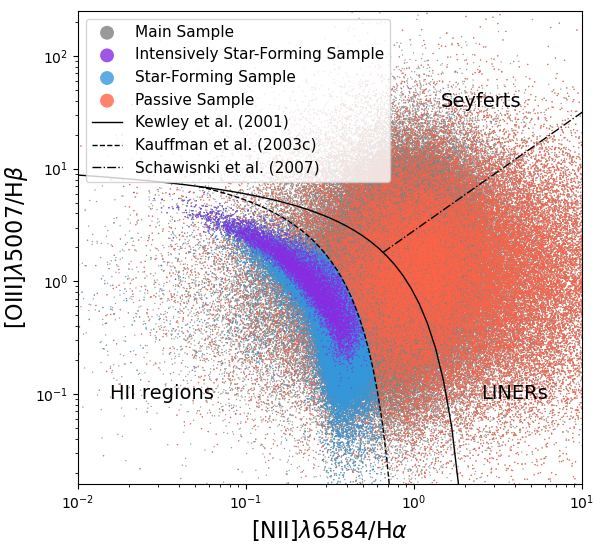}
\caption{Emission-line flux ratio [NII]$\lambda$6583/H$\alpha$ as a function of [OIII]$\lambda$5007/H$\beta$. Grey, violet, blue and red points represent the Main, Intensively Star-Forming, Star-Forming and Passive Samples, respectively. Full, dashed and dot-dashed black lines represent the demarcation lines proposed by \cite{Kewley_etal_2001}, \cite{Kauffmann_etal_2003c} and \cite{Schawinski_etal_2007}, respectively.}
\label{Fig_-_BPT_selection}
\end{center}
\end{figure}

	The analysis of the results is based on four main samples of galaxies. 	The {Main Sample (MS)} is defined by a cut in apparent magnitude in the $r-$band of $14.5 \leq m_r \leq 17.77$ and in redshift of $0.04 \leq z \leq 0.4$. The first criterion takes into consideration photometric completeness of the survey, whereas the second aims to remove low-redshift interlopers while also assuring that the most prominent optical emission lines (e.g. H$\beta\lambda$4861; [OIII]$\lambda$5007; [OI]$\lambda$6300; H$\alpha\lambda$6563; [NII]$\lambda$6583; [SII]$\lambda\lambda$6717,6731) fall within the observed wavelength range. Duplicates where removed by selecting objects classified as galaxies (`$\mathtt{specClass} = 2$') from the `$\mathtt{SpecObj}$' table created by the survey, which lists fibres categorised by `$\mathtt{SciencePrimary}$' (i.e. primary observation of the object). These criteria lead to the selection of 613592 objects ($\sim$66\% of the DR7 galaxy sample).
	
	The {Star-Forming Sample (SFS)} is defined by the criteria of MS in combination with: $\mathrm{EW(H\alpha)} \geq 3$ \AA , signal-to-noise at the normalisation wavelength of $\mathrm{S/N}(\lambda_{0}) \geq 3$, and the \cite{Kauffmann_etal_2003c} demarcation line. Similar to the criteria adopted by \cite{Asari_etal_2007}, these conditions select for emission-line galaxies with relatively good spectral quality located in the `SF' locus in the [NII]$\lambda$6583/H$\alpha$ and [OIII]$\lambda$5007/H$\beta$ emission-line diagnostic diagram (\citealt{Baldwin_Phillips_Terlevich_1981}). This selects 195479 objects ($\sim$31.9\% of MS). Moreover, the	{Intensively Star-Forming Sample (ISFS)} is defined by the criteria of SFS with a cut of $\mathrm{EW(H\alpha)} \geq 75$ \AA. This isolates galaxies in which the nebular continuum contribution is particularly high. This selects 8051 objects ($\sim$1.3\% of MS). 

	Finally, the {Passive Sample (PS)} is defined by the criteria of MS in combination with: $\mathrm{EW(H\alpha)} \leq 0.5$ \AA\ and $\mathrm{S/N}(\lambda_{0}) \geq 3$. Similarly, \cite{Mateus_etal_2006} defined `passive' or `lineless' galaxies based on \EWha\ and $\mathrm{EW(H\beta)} \leq 1$ \AA, whereas \cite{CidFernandes_etal_2011} defined `passive' as galaxies with \EWha\ and $\mathrm{EW(NII)} \leq 0.5$. This selects 103510 objects ($\sim$16.9\% of MS).

	Figure \ref{Fig_-_SN_F_EW} shows the distributions for the different samples of the signal-to-noise at the normalisation wavelength S/N(4020 \AA), the flux of the H$\alpha$ emission line F(H$\alpha$) and the H$\alpha$ equivalent width \EWha. Moreover, Figure \ref{Fig_-_BPT_selection} displays the location of the four samples in the \cite{Baldwin_Phillips_Terlevich_1981} diagnostic diagram, whereas Figures \ref{Fig_-_MS_-_fit_example_1}--\ref{Fig_-_PS_-_fit_example_2} show \Fado\ and \SL\ fit results for two randomly selected galaxies from each sample in the same format adopted in Figure \ref{Fig_-_text_fit_example}. 
		
\subsection{Total stellar mass, mean age and mean metallicity distributions}\label{SubSection_-_MtZ_Distributions}

\begin{figure*}[!t] 
\begin{center}	
\includegraphics[width=\textwidth]{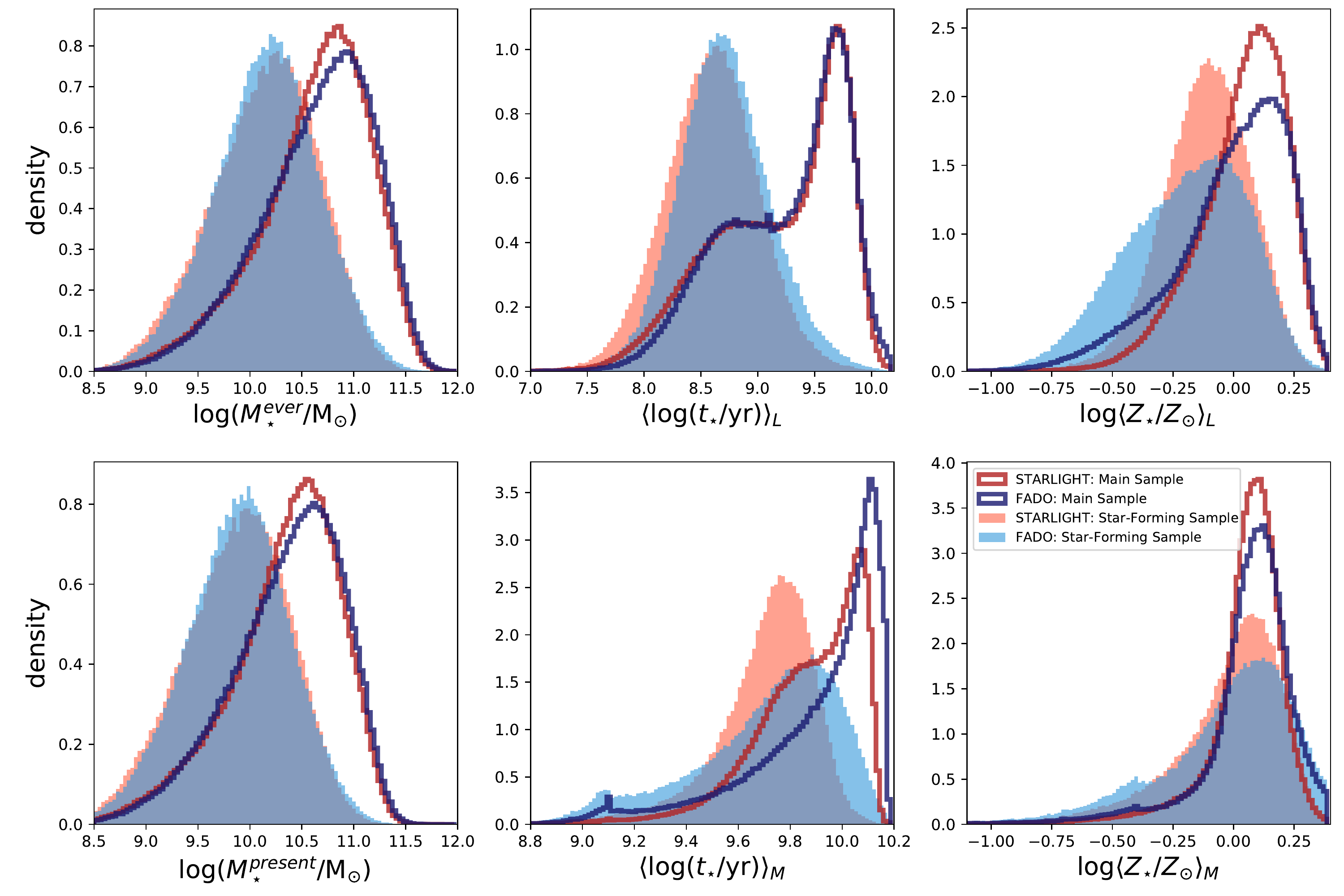}
\caption{Density distributions of the total stellar mass $M_\star$ in solar units $M_{\odot}$ (left-hand side panels), mean stellar age $\langle \log t_\star \rangle$ in years (centred panels) and mean stellar metallicity $\log \langle Z_\star \rangle$ in solar units (right-hand side panels). Blue and red colours represent results from \Fado\ and \SL, respectively, for the Main Sample (darker colours) and Star-Forming Sample (lighter colours). Top and bottom mass panels display, respectively, the  total mass ever-formed throughout the life of the galaxy $M_\star^{ever}$ and the presently available total stellar mass $M_\star^{present}$. Moreover, top and bottom age and metallicity panels represent their light- and mass-weighted versions, respectively.}
\label{Fig_-_MtZ_Histograms_-_MS_SFS}
\end{center}
\end{figure*}
\begin{figure*}[!t] 
\begin{center}	
\includegraphics[width=\textwidth]{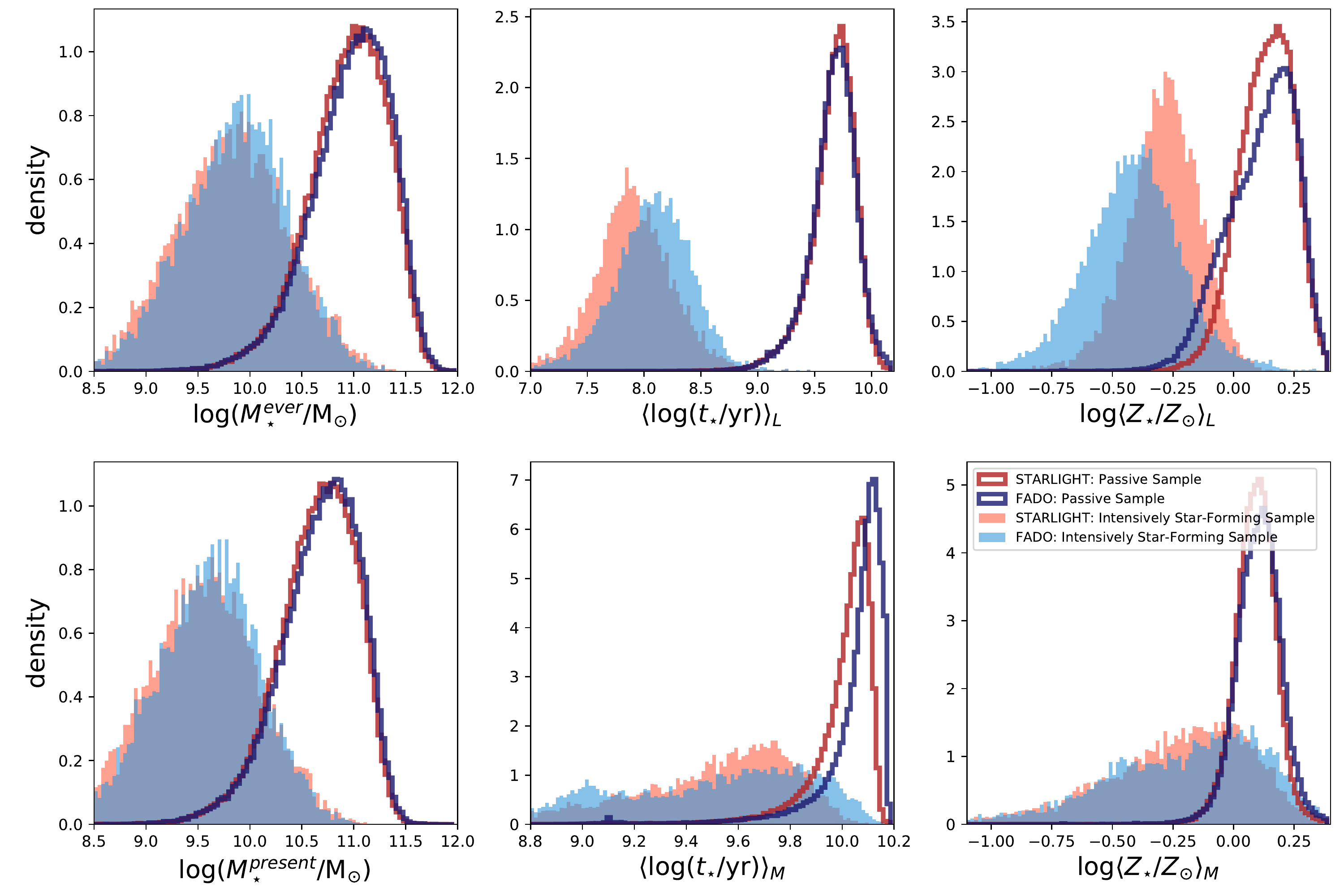}
\caption{Density distributions of the total stellar mass $M_\star$ in solar units $M_{\odot}$ (left-hand side panels), mean stellar age $\langle \log t_\star \rangle$ in years (centred panels) and mean stellar metallicity $\log \langle Z_\star \rangle$ in solar units (right-hand side panels). Blue and red colours represent results from \Fado\ and \SL, respectively, for the Passive Sample (darker colours) and Intensively Star-Forming Sample (lighter colours). Other legend details are identical to those in Fig. \ref{Fig_-_MtZ_Histograms_-_MS_SFS}.}
\label{Fig_-_MtZ_Histograms_-_PS_ISFS}
\end{center}
\end{figure*}

	Figure \ref{Fig_-_MtZ_Histograms_-_MS_SFS} shows on the left-hand side panels the distributions of the total stellar mass ever-formed $M_\star^{ever}$ (top) and presently available $M_\star^{present}$ (bottom) (i.e. after correcting for mass loss through regular stellar evolution) for the MS and SFS galaxy populations. Results show that both codes yield very similar mass distributions. Indeed, the difference between the average total stellar mass from \SL\ and \Fado\ is $\sim$0.02 and 0.01 dex for both masses when it comes to the MS and SFS, respectively. Figure \ref{Fig_-_MtZ_Histograms_-_PS_ISFS} compares the ISFS and PS distributions and shows similar trends, with the difference between the average total stellar mass from \SL\ and \Fado\ being $\sim$0.04 and 0.03 dex, respectively.

	The age and  metallicity distributions of the stellar populations contributing to the best-fit solution can be summarised in terms of their mean values, which can be interpreted as a first order moment of the overall stellar age and metallicity of a given galaxy. Following \cite{CidFernandes_etal_2005}, the mean logarithmic stellar age weighted by light or mass can be defined as, respectively,
	
	\begin{equation}
	\langle \log t_\star \rangle _L = \sum_{i=1}^{N_\star} \gamma_i \cdot \log t_i ,
																				\end{equation}
	\begin{equation}	
		\langle \log t_\star \rangle _M = \sum_{i=1}^{N_\star} \mu_i \cdot \log t_i ,
																				\end{equation}
																				
\noindent	where $t_i$ is the age of the $i^{\mathrm{th}}$ stellar element (i.e. SSP, in this case) in the base library, and $\gamma_i$ and $\mu_i$ are its light and mass relative contributions, respectively. The term $\mu_i$ refers to the mass fraction that takes into consideration the amount of stellar matter returned to the interstellar medium through stellar evolution, thus being  related to $M_{\star}^{present}$. 

	Congruously, assuming that $Z_i$ represents the relative metallicity contribution of the $i^{\mathrm{th}}$ SSP to the best-fit solution, then the mean logarithmic stellar metallicity weighted by light and mass can be defined as, respectively, 

	\begin{equation}
	 \log \langle Z_\star \rangle _L = \log \sum_{i=1}^{N_\star} \gamma_i \cdot  Z_i ,
																				\end{equation}
	\begin{equation}	
	 \log \langle Z_\star \rangle _M = \log \sum_{i=1}^{N_\star} \mu_i \cdot Z_i .
																				\end{equation}
										
	Applying these definitions, Figure \ref{Fig_-_MtZ_Histograms_-_MS_SFS} shows the distribution of the mean stellar age $\langle \log t_\star \rangle$ (centred panels) and mean stellar metallicity $\log \langle Z_\star \rangle$ (right-hand side panels) estimated with \Fado\ (blue) and \SL\ (red).  The MS distributions show that, overall, the \Fado\ and \SL\ results are once more rather similar, with a few noticeable differences: (a) the metallicity distributions of \Fado\ are broader than that of \SL, (b) the $\langle \log t_\star \rangle_M$ absolute maxima differ by $\sim$0.1 dex between \Fado\ and \SL, and (c) the $\log \langle  Z_\star \rangle_L$ and $\log \langle  Z_\star \rangle_M$ absolute maxima differ by $\sim$0.1 dex between codes. Results for the SFS are relatively clearer: (a) the absolute maxima in the $\langle \log t_\star \rangle_L$ differ by $\sim$0.1 dex between \Fado\ and \SL, with the former estimating slightly higher mean ages, (b) \Fado\ displays a broader $\langle \log t_\star \rangle_M$ distribution than \SL, and (c) \Fado\ also displays broader metallicity distributions than \SL, with \SL\ preferring values closer to solar metallicity. 
	
	It is worth noting that the overall MS mass and age distributions from \SL\ illustrated in Figure \ref{Fig_-_MtZ_Histograms_-_MS_SFS} are qualitatively in agreement with \cite{Mateus_etal_2006}, which analysed SDSS DR 7 with \SL. Moreover, the broadening of the age and metallicity distributions going from \SL\ to \Fado\ in MS is also qualitatively similar to that reported by \cite{Werle_etal_2020} when going from purely-optical spectroscopy to UV photometry plus optical spectroscopy using \SL, even though that work is focussed on early-type galaxies.

	Comparing results from ISFS and PS in Figure \ref{Fig_-_MtZ_Histograms_-_PS_ISFS} shows that: (a) \Fado\ and \SL\ mean age and metallicity distributions for the PS are very similar, with the main difference being the mass-weighted age distribution of \Fado\ being narrower and with higher maxima in relation to \SL, (b) \SL\ light-weighted age (metallicity) distribution peaks at a lower (higher) value than \Fado\ when it comes the ISFS, and (c) mass-weighted age and metallicity ISFS distributions in both codes are much wider than their light-weighted counterparts.

	Interestingly, the light-weighted age distributions for MS in Figure \ref{Fig_-_MtZ_Histograms_-_MS_SFS} show two peaks around $\sim$8.7 and 9.7 dex for both codes, which seem to visually match their SFS and PS distributions, respectively. This could be interpreted at first glance as evidence for the bimodal distribution of galaxies (e.g. \citealt{Gladders_Yee_2000, Strateva_etal_2001, Blanton_etal_2003, Kauffmann_etal_2003b, Baldry_etal_2004, Mateus_etal_2006}).  Furthermore, the MS mass-weighted age distribution from \SL\ displays two peaks around $\sim$9.8 and 10.1 dex which, once again, match its SFS and PS distributions.  However, the MS mass-weighted age distribution from \Fado\ shows only one prominent peak around $\sim$10.1 dex, which is congruent with its PS distribution. In addition, the SFS mass-weighted age distribution from \Fado\ (a) is much broader than that of \SL\ and (b) displays a conspicuous relative maximum at $\sim$9.1 dex of unknown origin, also present in the MS distribution. 
	
	At least two factors could be behind the differences between the codes regarding these particular features: (a) the different mathematical (e.g. noise treatment and minimisation procedure) and physical approaches of the two codes (e.g. nebular continuum treatment) and (b) the intrinsic degeneracies between the SSPs in the adopted library (e.g. \citealt{Faber_1972, Worthey_1994, CidFernandes_etal_2005}). In fact, it is worth considering that CGP19 found that nebular contamination in synthetic SF galaxies leads \SL\ to favour best-fit solutions where most of the light comes from a combination of very young and very old SSPs (i.e. bimodal best-fit solutions). A dedicated study exploring, for instance, how variations of the SSPs in the stellar library (e.g. \citealt{Leitherer_etal_1999, Bruzual_Charlot_2003, SanchezBlazquez_etal_2006, Molla_GarciaVargas_Bressan_2009}) impact the distributions of these stellar properties would help to clarify the sources behind the MS distribution features observed in each code.

	To put the results of Figures \ref{Fig_-_MtZ_Histograms_-_MS_SFS} and \ref{Fig_-_MtZ_Histograms_-_PS_ISFS} into context, it is also worth noting that \cite{CidFernandes_etal_2005} showed that the original version of \SL\ can recover the stellar mass, mean age and mean metallicity within $\sim$0.1 and $\sim$0.2 dex when modelling purely stellar synthetic spectra with S/N$\sim$10. Later, \cite{CidFernandes_etal_2014} used simulations of CALIFA data (\citealt{Sanchez_etal_2012}) and found that the same estimated stellar properties can be affected by uncertainties between $\sim$0.1 and 0.15 dex related to noise alone and $\sim$0.15 and 0.25 dex when it comes to changes in the base models.  Similar accuracy values were found in CGP19 regarding both \Fado\ and \SL. Among other factors, these biases can be attributed to the adopted stellar ingredients not adequately representing the wide range of stellar types and their evolutionary stages (e.g. \citealt{GonzalezDelgado_CidFernandes_2010, Ge_etal_2019}) or possible variations on the stellar initial mass function (e.g. \citealt{Conroy_Gunn_White_2009, Fontanot_2014, Barber_Schaye_Crain_2019}).

\begin{table*}[]
\begin{center}	
\caption{Average stellar properties of the galaxy samples.}
\begin{tabular}{ccccccccc}
  & \multicolumn{2}{c|}{Main Sample} & \multicolumn{2}{c|}{Star-Forming Sample} & \multicolumn{2}{c|}{Intensively SF Sample} & \multicolumn{2}{c}{Passive Sample}\\ \cline{2-9}
  
$\mu\pm\sigma$  & \Fado & \multicolumn{1}{c|}{\SL} & \Fado & \multicolumn{1}{c|}{\SL} & \Fado & \multicolumn{1}{c|}{\SL} & \Fado & \SL\\ \cline{1-9}

${\log M_{\star}^{ever}}$ & 10.64$\pm$0.55 & \multicolumn{1}{c|}{10.62$\pm$0.54} & 10.15$\pm$0.51 & \multicolumn{1}{c|}{10.14$\pm$0.52} & 9.85$\pm$0.51 & \multicolumn{1}{c|}{9.81$\pm$0.54} & 10.97$\pm$0.39 & 10.94$\pm$0.38\\

${\log M_{\star}^{present}}$ & 10.35$\pm$0.54 & \multicolumn{1}{c|}{10.33$\pm$0.53} & 9.87$\pm$0.50 & \multicolumn{1}{c|}{9.86$\pm$0.51} & 9.59$\pm$0.50 & \multicolumn{1}{c|}{9.55$\pm$0.52} & 10.68$\pm$0.38 & 10.65$\pm$0.38\\

${\langle \log t_\star \rangle_L}$ & 9.24$\pm$0.53 & \multicolumn{1}{c|}{9.18$\pm$0.57} & 8.71$\pm$0.41 & \multicolumn{1}{c|}{8.59$\pm$0.41} & 8.07$\pm$0.35 & \multicolumn{1}{c|}{7.90$\pm$0.36} & 9.66$\pm$0.22 & 9.66$\pm$0.21\\

${\langle \log t_\star \rangle_M}$ & 9.89$\pm$0.28 & \multicolumn{1}{c|}{9.87$\pm$0.22} & 9.70$\pm$0.30 & \multicolumn{1}{c|}{9.70$\pm$0.21} & 9.47$\pm$0.43 & \multicolumn{1}{c|}{9.47$\pm$0.36} & 10.05$\pm$0.15 & 10.00$\pm$0.13\\



${\log \langle Z_\star \rangle_L}$ & -0.01$\pm$0.24 & \multicolumn{1}{c|}{0.03$\pm$0.18} & -0.20$\pm$0.26 & \multicolumn{1}{c|}{-0.12$\pm$0.19} & -0.42$\pm$0.21 & \multicolumn{1}{c|}{-0.29$\pm$0.16} & 0.12$\pm$0.14 & 0.13$\pm$0.11\\

${\log \langle Z_\star \rangle_M}$ & 0.06$\pm$0.21 & \multicolumn{1}{c|}{0.05$\pm$0.17} & -0.05$\pm$0.30 & \multicolumn{1}{c|}{-0.04$\pm$0.24} & -0.22$\pm$0.36 & \multicolumn{1}{c|}{-0.24$\pm$0.32} & 0.10$\pm$0.10& 0.09$\pm$0.09

\end{tabular}

\end{center}
{ \textbf{Notes.} Average values $\mu$ and corresponding standard deviations $\sigma$ of the total stellar mass (in solar masses $M_{\odot}$) ever-formed $M_{\star}^{ever}$ and presently available $M_{\star}^{present}$, mean stellar age (in years) weighted by light $\langle \log t_\star \rangle_L$ and mass $\langle \log t_\star \rangle_M$, and mean stellar metallicity (in solar units $Z_{\odot}=0.02$) weighted by light $\log \langle Z_\star \rangle_L$ and mass $\log \langle Z_\star \rangle_M$ for the Main, Star-Forming, Intensively Star-Forming and Passive Samples, as defined in Subsection \ref{SubSection_-_Sample_Selection}. }
\label{Table_-_MS_and_SFS}
\end{table*}
	
	The average MS, SFS, ISFS and PS population values for mass, age and metallicity are gathered in Table \ref{Table_-_MS_and_SFS} for each code. The differences between codes for total mass, mean age and mean metallicity can go up to $\sim$0.06 for MS and $\sim$0.12 dex for SFS, respectively. The most significant difference in MS relates to the average light-weighted age, whereas for SFS the most prominent discrepancies between codes rest on the light-weighted age and metallicity and amount to $\sim$0.12 and 0.08 dex, respectively. Moreover, the ISFS and PS results in this table show stellar properties differences between codes in the ranges of $\sim$0.02--0.17 dex for ISFS and $\sim$0--0.05 dex for PS. Although the main discrepancy in ISFS falls again on the average light-weighted age ($\sim$0.17 dex), the main difference between codes for PS has to do with the average mass-weighted age ($\sim$0.05 dex). The possible origins for these differences are explored in Subsection \ref{SubSection_-_Light_and_Mass_Distributions}. Finally, the working definition for the mass for the remainder of this work is that of $M_\star^{present}$.
		
\subsection{Light and mass distributions}\label{SubSection_-_Light_and_Mass_Distributions}	
	
	As the previous results have shown, the mean age $\langle \log t_\star \rangle$ and mean metallicity  $\log \langle  Z_\star \rangle$ parameters are powerful tools to evaluate the stellar properties of galaxies, particularly when grouped in populations of the same type. However, these parameters also lack the detailed information required to understand some of the differences between \Fado\ and \SL\ reported in the previous subsection. The same would be true for any ad-hoc binning schema that can be adopted to downscale the best-fit population vector (PV) into more easily manageable parameters. Indeed, a better strategy to study the code differences for each galaxy population is to look into the full information encoded in the best-fit PVs regarding the light and mass contributions of each stellar element. 

\begin{figure*}[!t] 
\begin{center}	
\includegraphics[width=0.9\textwidth]{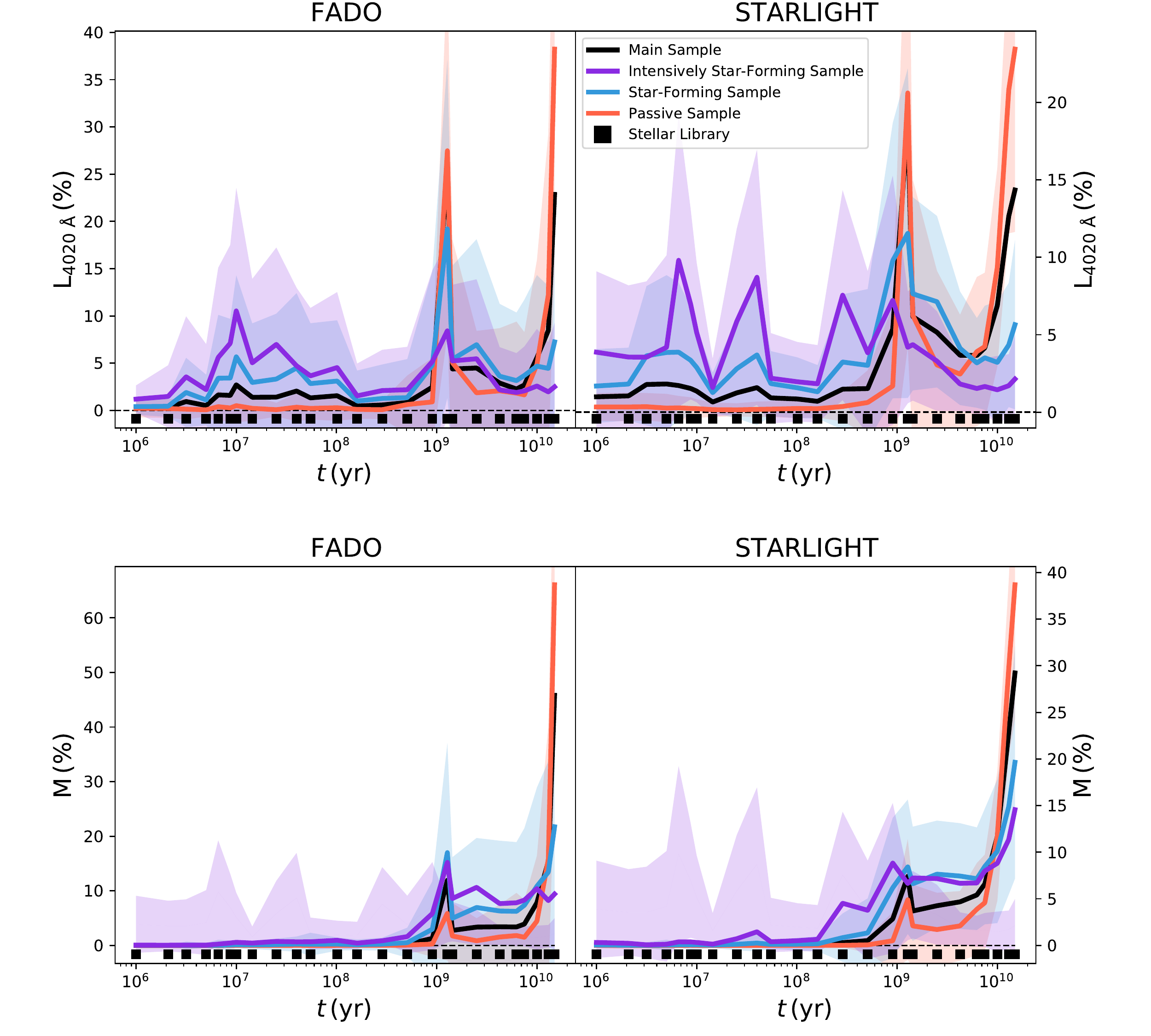}
\caption{ Light (top panels) and mass (bottom panels) relative contributions of the stellar library elements as a function of their age. Black, violet, blue and red lines represent the average results for the MS, ISFS, SFS and PS galaxy populations, respectively, while the surrounding shading  denotes their corresponding standard deviation. Black squares represent the age coverage of the adopted stellar library, with the six metallicity values summed at each age step.}
\label{Fig_-_SFH_vs_age}
\end{center}
\end{figure*}

	With this in mind, Figure \ref{Fig_-_SFH_vs_age} shows the light (top panels) and mass relative contributions (bottom panels) of each stellar element (i.e. SSP) in the adopted base library as a function age.  The relative contributions of the six metallicities in the  base are summed up along each age step (represented by the black squares). These results can be viewed as the light and mass  SFHs of each galaxy population, with Figure \ref{Fig_-_SFH_vs_age_-_cumulative} showing their cumulative versions.
	
	Several results are worth discussing. Firstly, the relative light contributions of young-to-intermediate SSPs with $t \! < \! 10^8$ yr increases with increasing \EWha\ in both codes, following the PS$\rightarrow$MS$\rightarrow$SFS$\rightarrow$ISFS sequence. The opposite trend is seen in old SSPs with $t \! > \! 10^9$ yr. This is to be expected, since young bright stellar populations dominate their older counterparts in terms of light output in ISFS and SFS galaxies, whereas the bulk of stellar content in PS is rather old and considerably less luminous. 
	Secondly, light distribution results show a peak around $t \! \sim \! 10^9$ yr in both codes, ranging between $\sim$5 (ISFS) and 25\% (PS). \cite{Asari_etal_2007} reported a similar hump around 1 Gyr in the SFR when displayed as a function of stellar age.  The authors carried tests with \SL\ and found that hump disappears after changing the stellar library of SSPs from STELIB (\citealt{Bruzual_Charlot_2003}), as in this work, to MILES (\citealt{SanchezBlazquez_etal_2006}).
		
	The comparison between \Fado\ and \SL\ is particularly revealing when it comes to the relative contribution of young and intermediate aged SSPs.  Results show that SSPs with ages $t \! < \! 10^8$ yr  contribute with more light in \SL\ than in \Fado, which is increasingly noticeable with increasing \EWha\ (PS$\rightarrow$MS$\rightarrow$SFS$\rightarrow$ISFS). Indeed, Figure \ref{Fig_-_SFH_vs_age_-_cumulative} shows that the light contribution in \SL\ is greater by 5.41\% than in \Fado\ at $t \! = \! 10^7$ yr for the ISFS, reaching 9.11\% around $t \! = \! 10^9$ yr. This means that \SL\ overestimates the relative light contribution of young stellar populations in relation to \Fado\ (or vice versa), thus accounting for the $\sim$0.12 and 0.17 dex difference in $\langle \log t_\star \rangle_L$ between the codes for the SFS and ISFS populations, respectively (Table \ref{Table_-_MS_and_SFS}). 
	
	When it comes to the mass distributions, the differences between the codes are considerably less pronounced. The main reason for this lies in the fact that, in the local universe, most stellar mass is locked into older stellar populations. This is best exemplified by the increasing importance of $t \! > \! 10^{10}$ yr SSPs with decreasing \EWha, following the ISFS$\rightarrow$SFS$\rightarrow$MS$\rightarrow$PS sequence. Apart from the peak around $t \! \sim \! 10^9$ yr already noted, it is worth observing that \SL\ results in the ISFS show higher mass contributions from SSPs at $t \sim 3 \cdot 10^8$ yr than \Fado. This seems to be offset by an excess of $\sim$5\% of SSPs with $t > 10^{10}$ yr, which could explain why the average mean stellar age of \SL\ for the SFS ($\sim$9.7 dex) and ISFS ($\sim$9.47 dex) documented in Table \ref{Table_-_MS_and_SFS} is serendipitously identical to that of \Fado. In general, these differences between \Fado\ and \SL\ when it comes to both the light and mass distributions are somewhat similar to those reported by \cite{Werle_etal_2019} when adopting a new version of \SL\ that incorporates UV photometry.

\begin{figure*}[!t] 
\begin{center}	
\includegraphics[width=0.9\textwidth]{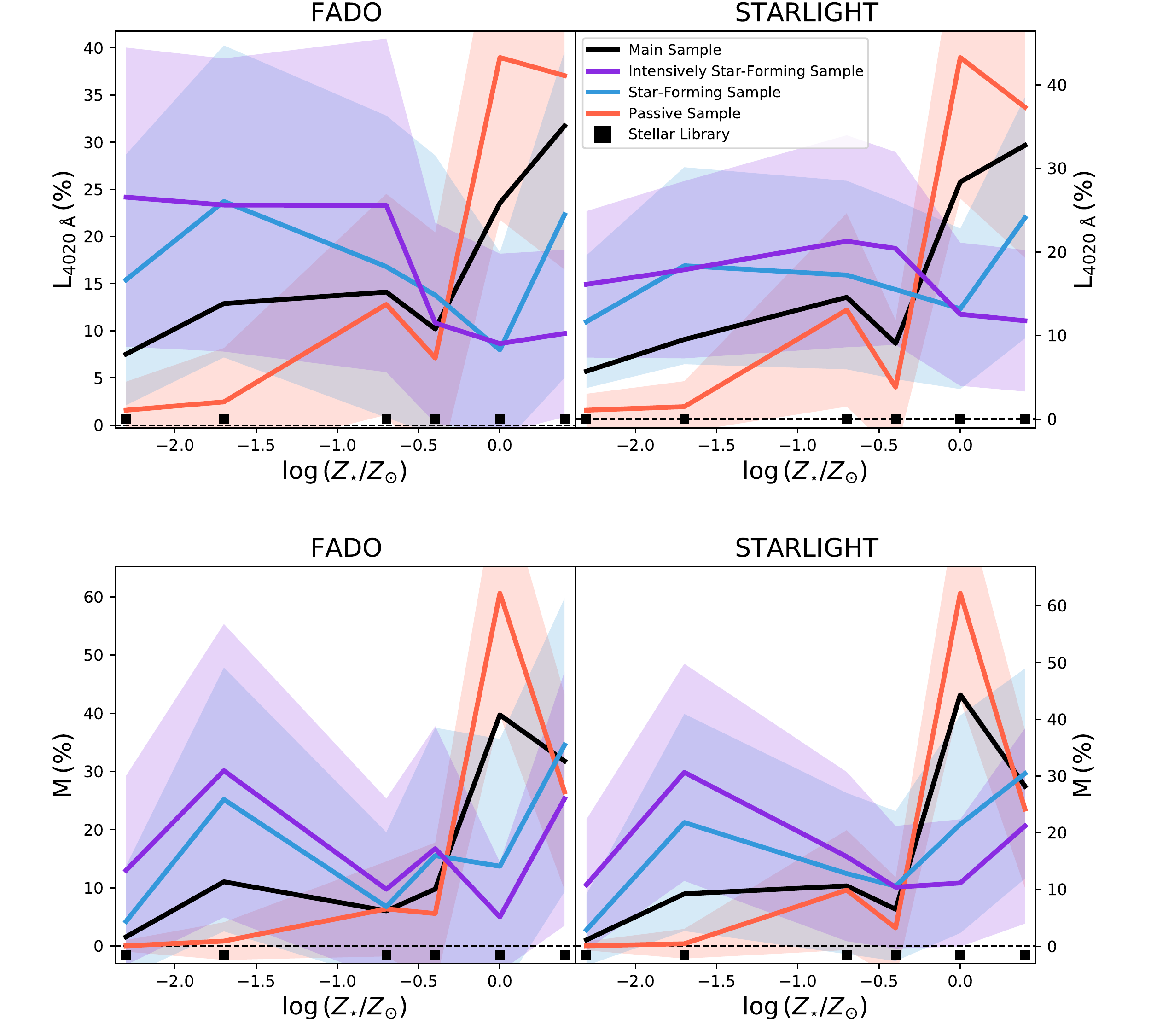}
\caption{ Light (top panels) and mass (bottom panels) relative contributions of the stellar library elements as a function of their metallicity. Black, violet, blue and red lines represent the average results for the MS, ISFS, SFS and PS galaxy populations, respectively, while the surrounding shading  denotes the their corresponding standard deviation. Black squares represent the metallicity coverage of the adopted stellar library, with the 25 age values summed at each metallicity step.}
\label{Fig_-_SFH_vs_Z}
\end{center}
\end{figure*}

	Exploring further the information encoded in the PVs, Figure \ref{Fig_-_SFH_vs_Z} shows the light (top panels) and mass relative contributions (bottom panels) of each stellar element in base as a function metallicity.  In contrast to Figure \ref{Fig_-_SFH_vs_age}, the relative contributions of the 25 age steps in the base are now summed along the six metallicities in the base library (represented by the black squares). Other remaining plot details are similar to those of Figure \ref{Fig_-_SFH_vs_age}. 
	
	The light distributions displayed on the top panels show for both codes that the relative contributions of the lowest metallicities increase with increasing \EWha\ following PS$\rightarrow$MS$\rightarrow$SFS$\rightarrow$ISFS, with the opposite trend at the highest metallicities. The inversion point occurs somewhere between $2 Z_\odot / 5$ and $Z_\odot$ (i.e. $\log ( Z_\star/Z_\odot) = -0.398$ and 0, respectively). Moreover, the best-fit solutions from \Fado\ have higher light contributions from the two lowest metallicities than \SL, with SSPs with $Z_\odot / 200$ and $Z_\odot / 50$ (i.e. $\log ( Z_\star/Z_\odot) \simeq -2.301$ and -1.699, respectively) contributing $\sim$20-25\% in \Fado\ and $\sim$15-20\% in \SL. This is best displayed in Figure \ref{Fig_-_SFH_vs_Z_-_cumulative}, which shows that the difference in the relative contribution between \Fado\ and \SL\ is already of 8.05\% at $Z_\odot / 200$ and reaches 13.51\% by $Z_\odot / 50$ for ISFS. This indicates that \SL\ overestimates the metallicity of SF galaxies in comparison to \Fado\ (or vice versa). However, this interpretation is strongly tempered by the particularly large standard deviations observed for SFS and ISFS populations. 
	
	As a side note, CGP19 found using synthetic galaxy spectra that \SL\ systematically underestimates the light-weighted mean meallicity by up to $\sim$0.6 dex with decreasing age of the galaxy (i.e. increasing \EWha) for $t<10^9$ yr, whereas the mass-weighted can be underestimated for $10^7<t<10^9$ yr by up to $\sim$0.6 dex or overestimated by up to $\sim$0.4 dex for $t<10^7$ yr. Similar results where found in P21. Although the details of these trends depend on the SFH of the models (e.g. instantaneous or continuous), adopted fitting methodology (e.g. including or excluding the emission lines and the Balmer and Paschen discontinuities) and S/N, the important fact to bear in mind is that these tests were carried out assuming a constant solar stellar metallicity of $Z_\odot=0.02$. Therefore, metallicity results detailed in Figures \ref{Fig_-_MtZ_Histograms_-_MS_SFS} and \ref{Fig_-_SFH_vs_Z} cannot be easily compared to the results of CGP19 or P21. In contrast, the overall mean age trends documented in this work are compatible to those found in CGP19 or P21.
	
	Finally, the bottom panels of Figure \ref{Fig_-_SFH_vs_Z} shows that mass distributions differ very little between codes, in a similar vein to those in Figure \ref{Fig_-_SFH_vs_age}. However, it is interesting to note the negligible contribution of two lowest metallicities to the mass when it comes to the PS galaxies, in contrast to both SFS and ISFS. Coupled with the trends observed in the mass distributions as a function age, these results  point again to the idea of a bimodal distribution of galaxies (e.g. \citealt{Gladders_Yee_2000, Strateva_etal_2001, Blanton_etal_2003, Kauffmann_etal_2003b, Baldry_etal_2004}).

\subsection{Relations between mass, mean age and mean metallicity in star-forming galaxies}\label{SubSection_-_MtZ_Relations}	

\begin{figure*}[!t] 
\begin{center}	
\includegraphics[width=\textwidth]{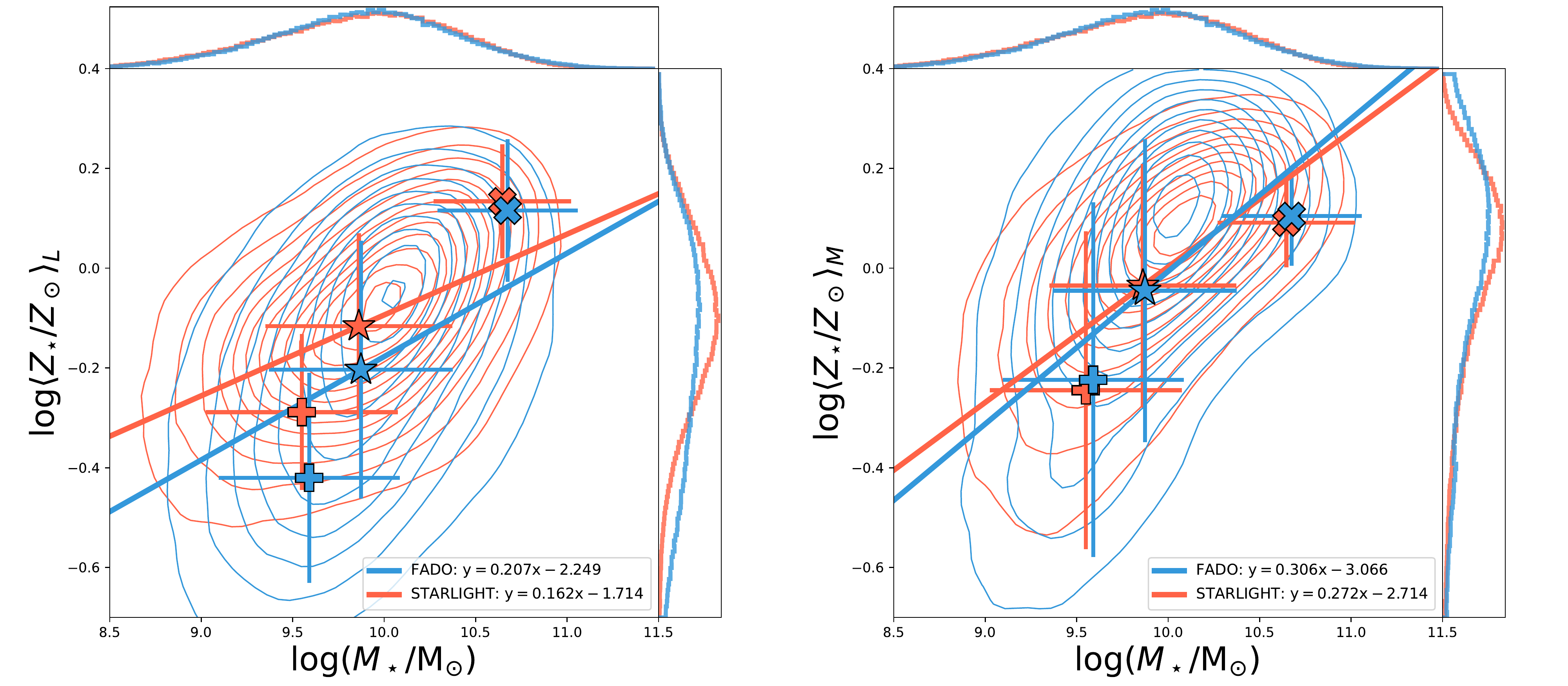}
\caption{ Light (left-hand side panels) and mass-weighted (right-hand side panels) mean stellar metallicity $\log \langle Z_\star \rangle$ as a function of total stellar mass $M_\star$ for the Star-Forming Sample. Blue and red lines and points represent \Fado\ and \SL\ results, respectively. The `\ding{58}', `$\bigstar$' and `\ding{54}' symbols represent the average values for the ISFS, SFS and PS populations, respectively, with standard deviation errorbars. Linear regression results for the SFS population from each code are present in the legend of each plot.}
\label{Fig_-_SFS_-_M_vs_Z}
\end{center}
\end{figure*}
\begin{figure*}[!t] 
\begin{center}	
\includegraphics[width=\textwidth]{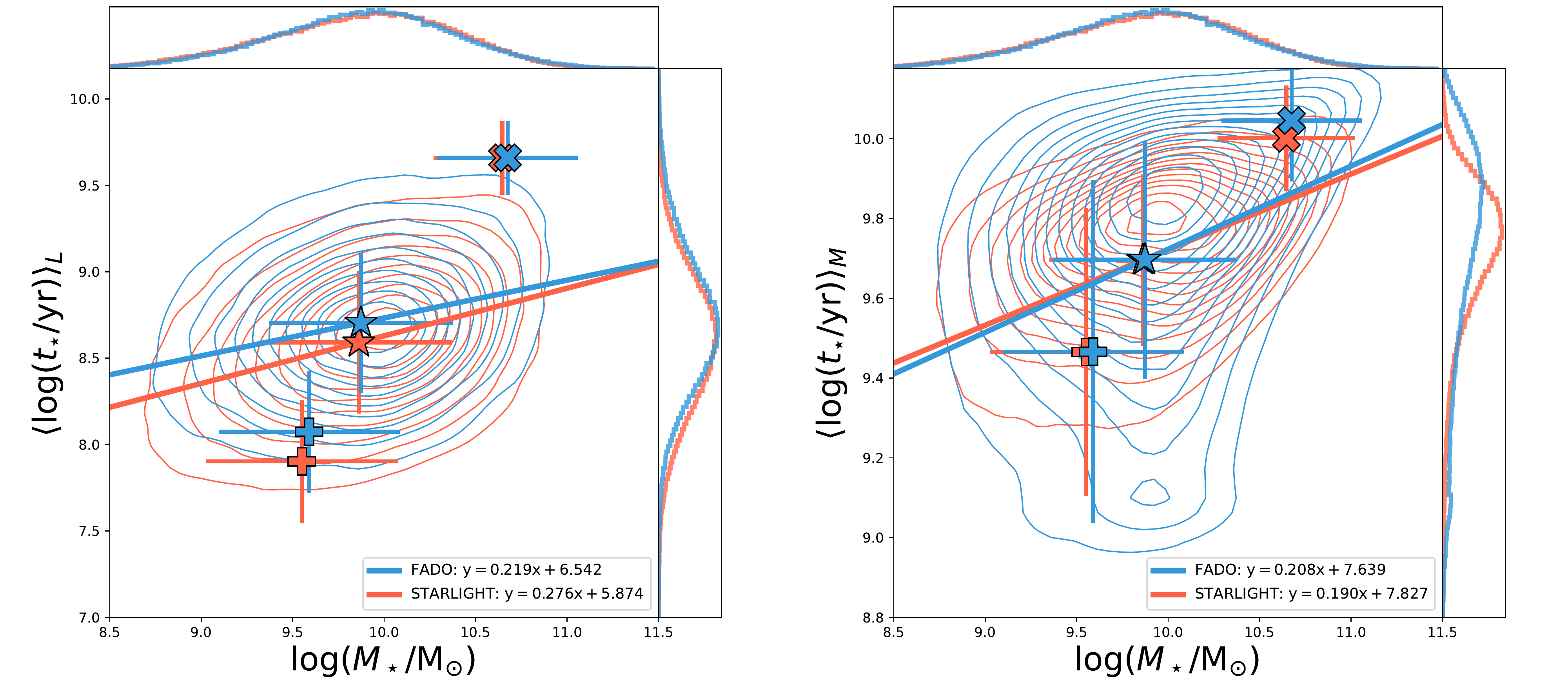}
\caption{Light (left-hand side panels) and mass-weighted (right-hand side panels) mean stellar age $\langle \log t_\star \rangle$ as a function of total stellar mass $M_\star$ for the Star-forming Sample. Other legend details are identical to those in Fig. \ref{Fig_-_SFS_-_M_vs_Z}.}
\label{Fig_-_SFS_-_M_vs_t}
\end{center}
\end{figure*}
\begin{figure*}[!t] 
\begin{center}	
\includegraphics[width=\textwidth]{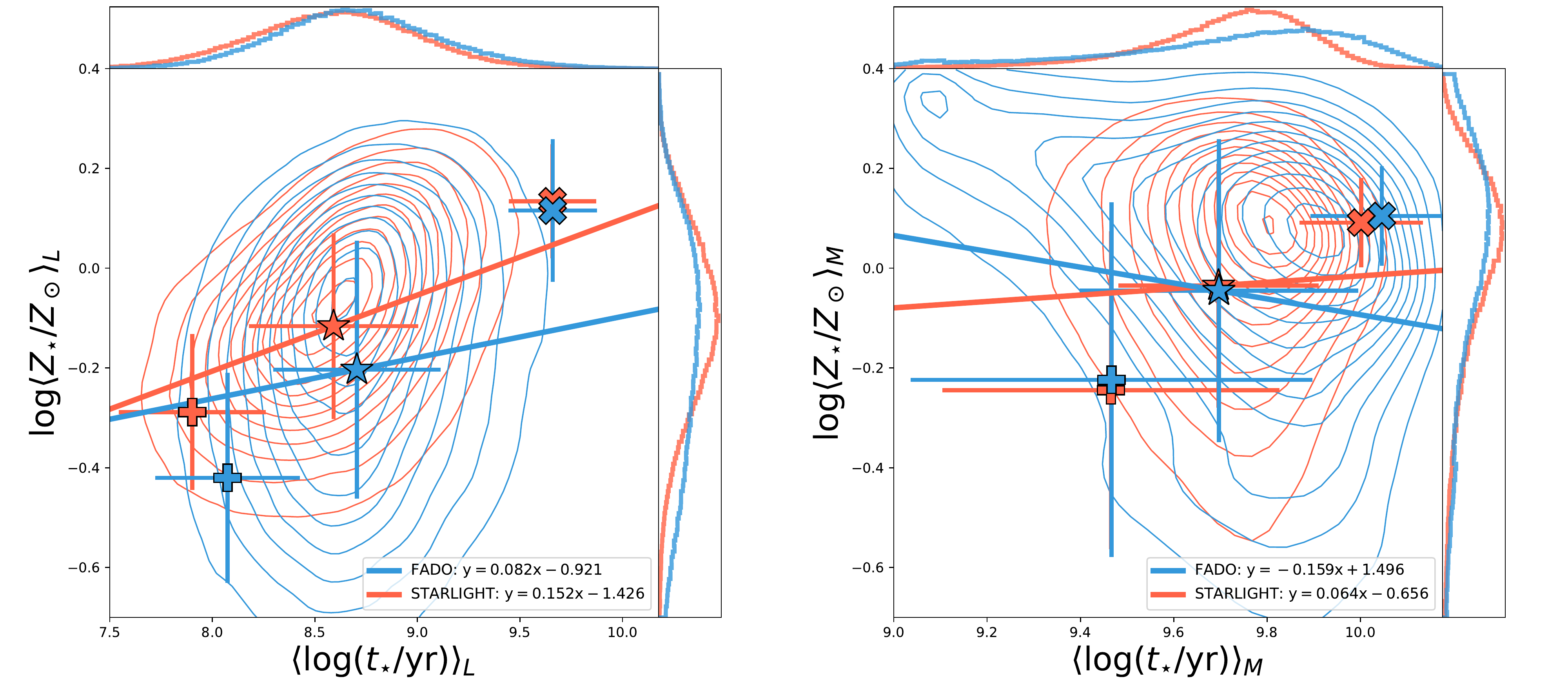}
\caption{ Light (left-hand side panels) and mass-weighted (right-hand side panels) mean stellar metallicity $\log \langle Z_\star \rangle$ as a function of mean stellar age $\langle \log t_\star \rangle$ for the Star-Forming Sample. Other legend details are identical to those in Fig. \ref{Fig_-_SFS_-_M_vs_Z}.}
\label{Fig_-_SFS_-_t_vs_Z}
\end{center}
\end{figure*}

	Figures \ref{Fig_-_SFS_-_M_vs_Z}, \ref{Fig_-_SFS_-_M_vs_t} and \ref{Fig_-_SFS_-_t_vs_Z} show in the contour lines the relations between the total stellar mass $M_\star$, the mean stellar age $\langle \log t_\star \rangle$ and the mean stellar metallicity $\log \langle  Z_\star \rangle$ for the SFS galaxy population (Figures \ref{Fig_-_MS_-_M_vs_Z}--\ref{Fig_-_MS_-_t_vs_Z} show similar plots for the MS), with special symbols representing average values for the ISFS, SFS and PS populations. 
	
	The mass-metallicity relation displayed in Figure \ref{Fig_-_SFS_-_M_vs_Z} shows an increasing divergence between the codes in light-weighted metallicity as \EWha\ increases and metallicity decreases. This is illustrated by the somewhat steeper linear regression of \Fado\ in comparison with \SL\ and the difference between the average values for ISFS, SFS and PS populations. In contrast, both codes show similar results when it comes to the mass-weighted metallicity. Both trends are in keeping with the discussion of Figure \ref{Fig_-_SFH_vs_Z}. 

	The mass-age relation in Figure \ref{Fig_-_SFS_-_M_vs_t} also shows interesting results. For instance, \SL\ shows a systematic light-weighted age underestimation in comparison with \Fado\ with increasing \EWha, as shown by the increasing divergence between codes following the sequence PS$\rightarrow$SFS$\rightarrow$ISFS. This is particularly interesting since the age distributions from both codes are rather similar for SFS. At the same time, \Fado\ shows a mass-weighted mean age distribution that is broader and with a lower absolute maximum when compared to \SL, even though the average mean stellar ages of the ISFS and SFS samples for each code are very similar, as noted in Table \ref{Table_-_MS_and_SFS} and in Subsection \ref{SubSection_-_MtZ_Distributions}. 

	Finally, the age-metallicity relation in Figure \ref{Fig_-_SFS_-_t_vs_Z} gives another perspective to the age and metallicity differences between codes seen in Figures \ref{Fig_-_SFS_-_M_vs_Z} and \ref{Fig_-_SFS_-_M_vs_t}. The contour lines alone clearly illustrate the wider age and metallicity distributions from \Fado\ in comparison with \SL. The mass-weighted linear regression lines indicate the overall SFS distributions are intrinsically different between codes, even though the average values for each population are rather similar. The \Fado\ results in Figure \ref{Fig_-_SFS_-_t_vs_Z} are particularly interesting in this regard, since light-weighted metallicity increases with age but its mass-weighted counterpart decreases with its corresponding age. Indeed, the distributions of $\langle \log t_\star \rangle_M$ and $\log \langle  Z_\star \rangle_M$ from \Fado\ are significantly broader that those from \SL\ (as illustrated in Figure \ref{Fig_-_MtZ_Histograms_-_MS_SFS}), with relative maxima at $\langle \log t_\star \! \rangle_M \! \sim 9.1$ yrs and $\log \langle  Z_\star \rangle_M \! \sim \! -2.1$, weighting towards an anti-correlation between age and metallicity. This anti-correlation can be attributed, at least partly, to the well-known age-metallicity degeneracy, with young metal-rich stellar populations being indistinguishable of old metal-poor populations from the point of view of spectral synthesis (e.g. \citealt{Faber_1972, OConnell_1980, Bressan_Chiosi_Tantalo_1996, Pelat_1997, Pelat_1998, CidFernandes_etal_2005}). Although this is especially noticeable when it comes to the SFS, the mass-weighted age-metallicity relation presented in Figure \ref{Fig_-_MS_-_t_vs_Z} suggests that this trend is also present in the MS.

\section{Discussion}\label{Section_-_Discussion}

\subsection{Impact of the nebular continuum}\label{Subsection_-_Impact_of_the_Nebular_Continuum}

	An important factor to consider in the interpretation of the results presented in Section \ref{Section_-_Results} is the potential impact of the nebular continuum modelling approach in each code (or lack thereof) has on the estimated stellar properties of SF galaxies. Moreover, one wonders if this effect can be distinguish from (a) the intrinsic uncertainties (e.g. adequacy of age and metallicity coverages) and degeneracies associated with the adopted physical ingredients (e.g. SSPs, extinction, and kinematics) and (b) the different mathematical methods adopted in each code (e.g. Metropolis algorithm coupled with simulated annealing in \SL\ and genetic differential evolution optimisation in \Fado).
	
	With this in mind, one of the main objectives of the methodology presented in Section \ref{Section_-_Methodology} was to focus on the impact of nebular continuum modelling in SF galaxies by reducing the number of variables in the code comparison, following a fitting strategy similar to that of CGP19+P21. However, there are important methodological differences between this study and CGP19+P21 that prevent a straightforward comparison between works, such as: (a) the synthetic galaxies analysed in CGP19+P21 have constant solar stellar metallicity, (b) the current work includes two extra sub-solar metallicities in the adopted stellar library, and (c) the most prominent emission lines were masked in \SL\ using individual spectral masks in this work, whereas CGP19+P21 adopted a general mask built from \SL\ tests using SDSS observations (\citealt{CidFernandes_etal_2005}).

	Notwithstanding, the question regarding the impact of the nebular continuum modelling approach on the inferred stellar properties remains. In order to address this issue from a different angle, \Fado\ was applied to the SFS and ISFS galaxies in a purely stellar mode (c.f. \citealt{Gomes_Papaderos_2017}) using the same input spectra and spectral fitting setup as in \SL . The objective is to model the spectra using only a combination of stellar components, similarly to \SL , and compare it with the previous \Fado\ results in which the stellar and nebular spectral continua were fitted self-consistently. 
	
	Figure \ref{Fig_-_MtZ_Histograms_-_PurelyStellar} compares the distributions of the stellar properties using \Fado\ in `purely stellar mode' ($\mathtt{ST}$mode) with the results presented in Section \ref{Section_-_Results} for \Fado\ in `full-consistency mode' ($\mathtt{FC}$mode) and \SL . Several interesting results are worth noting: (a) \FadoSTmode\ mass distributions are very similar to those of \FadoFCmode , (b) \FadoSTmode\  light-weighted mean age distribution is close to that of \SL\ for SFS and falls between \SL\ and \FadoFCmode\ for ISFS, and (c) \FadoSTmode\ mass-weighted metallicity distributions are slightly more skewed to higher values than \FadoFCmode\ for both SFS and ISFS, a trend similar to that observed in the \SL\ results. 
	
	On the one hand, the fact that the \FadoSTmode\ results are in general closer to \FadoFCmode\ than to \SL\ suggests that the code differences documented in Section \ref{Section_-_Results} are dominated by the fundamental mathematical and statistical differences between codes (e.g. different minimisation procedures), with different physical ingredients and methods (e.g. different nebular continuum modelling approaches and emission-line masking) playing a secondary role. However, it is reasonable to think that these two factors are likely interconnected, if not mutually dependent.
	
	On the other hand, the \FadoSTmode\ light-weighted age distributions for SFS and ISFS seem to skew from the \FadoFCmode\ distributions towards those of the \SL. This is more clearly illustrated in Figure \ref{Fig_-_SFH_vs_age_&_Z_-_PurelyStellar_-_cumulative}, which compares the light and mass distributions of the PVs of \FadoSTmode\ with those of \SL\ and \FadoFCmode . This figure shows that \FadoSTmode\ overestimates for ISFS the contribution of SSPs younger than $\leq 10^7$ yr ($10^9$) by $\sim$5.74\% (0.88\%) in relation to \FadoFCmode , while also overestimating the contribution of SSPs with metallicities $\leq Z_\odot / 200$ ($Z_\odot / 50$) by $\sim$9.68\% (8.7\%). These results show that the nebular continuum modelling approach impacts the \Fado\ results similarly to the light-weighted age trends presented in Subsection \ref{SubSection_-_Light_and_Mass_Distributions} and, therefore, indirectly suggest that the SFS \SL\ results are likely affected by the lack of a nebular continuum modelling recipe (even if its impact is mixed with other uncertainties inherent to population synthesis). One way to rigorously quantify this impact would be to apply \SL\ in `self-consistent mode' to these samples of SF galaxies, which obviously is not an option, even considering the code version first presented in \cite{LopezFernandez_etal_2016}.
	
	These results highlight an obvious yet elusive idea worth considering during the development of the next generation of spectral synthesis codes. The comparison of new codes, with evermore relevant physical ingredients (e.g. self-consistent dust or AGN treatment), with their older counterparts will necessarily be increasingly complex due to the introduction of more statistical uncertainties and degeneracies between the physical ingredients. Parallel tests with corresponding increasingly physical complexity using synthetic spectral data (e.g. CGP19; P21) will continue to be essential in such an endeavour.

\subsection{Potential implications}\label{Subsection_-_Implications}

	Assuming that purely stellar modelling in fact leads to an overestimation of mean stellar metallicity with decreasing mass or increasing \EWha, this means that the mass-metallicity relation presented in Figure \ref{Fig_-_SFS_-_M_vs_Z} could become  steeper with increasing redshifts as gas reservoirs are both increasingly more abundant and less chemically enriched. This raises the further doubt of how well the synthesis models can mimic the stellar content at large distances. Most commonly adopted evolutionary models are based on stellar libraries of stars in the solar vicinity (e.g. \citealt{Leitherer_etal_1999, Bruzual_Charlot_2003, Vazquez_Leitherer_2005, SanchezBlazquez_etal_2006, Molla_GarciaVargas_Bressan_2009, Rock_etal_2016}), which might not be representative of the stellar populations at high-redshifts, especially when it comes to extremely low metallicities and Population III stars (e.g. \citealt{Schaerer_2002}). The fact that AGN and nebular continua dilute absorption features (e.g. \citealt{Koski_1978, CidFernandes_StorchiBergmann_Schmitt_1998, Moultaka_Pelat_2000, Kauffmann_etal_2003c, Vega_etal_2009, Cardoso_Gomes_Papaderos_2016, Cardoso_Gomes_Papaderos_2017}) further exacerbates this problem from the point of view of spectral synthesis. However, this concern is counterbalanced with the increasingly sophisticated theoretical synthesis models (e.g. \citealt{Coelho_etal_2007, Molla_GarciaVargas_Bressan_2009, Leitherer_etal_2010, Coelho_2014, Stanway_Eldridge_2019, Coelho_Bruzual_Charlot_2020}) which can help bridge the gap towards better hybrid evolutionary models. 

	Moreover, a systematic mean stellar age underestimation when adopting a purely stellar modelling approach has repercussions for the current interpretation of the physical properties of young stellar populations. As noted by \cite{Reines_etal_2010}, estimating the physical properties (e.g. age and metallicity) of star clusters through population synthesis can only be reliably accomplished when modelling both the nebular continuum and emission-lines. The same is true for galaxies with relatively high specific SFRs in the local universe, such as dwarfs (e.g. \citealt{Papaderos_etal_1998, Papaderos_Ostlin_2012}) and `green peas' (e.g. \citealt{Izotov_Guseva_Thuan_2011, Amorin_etal_2012}), or at higher redshifts (e.g. \citealt{Zackrisson_Bergvall_Leitet_2008, Schaerer_deBarros_2009}). 
	
	From a different perspective, variations in the emission-line measurements could impact SFR estimations and, thus, indirectly affect estimations of the cosmic SF history (\citealt{Panter_etal_2007}).  The fact that \Fado\ measures emission-lines after subtracting the nebular continuum means that the EWs of the Balmer lines will always be greater than those based on a purely stellar modelling approach. At the same time, SFR estimators based on emission-line fluxes are not expected to change since the modelling (or not) of the nebular continuum does not impact the measured fluxes.

\section{Summary and conclusions}\label{Section_-_Conclusions}

	The population synthesis code \Fado\ (\citealt{Gomes_Papaderos_2017, Gomes_Papaderos_2018}) was applied to the main sample of galaxies from the SDSS (\citealt{York_etal_2000}) DR 7 (\citealt{Abazajian_etal_2009}) with the aim of re-evaluating the relations between the main stellar properties of galaxies (e.g. mass, mean age, and mean metallicity), particularly those of SF galaxies. The main reason for such re-analysis is the fact that \Fado\ is the first publicly available population synthesis tool to self-consistently model both the stellar and nebular continua. In fact, previous studies adopted purely stellar population synthesis codes (e.g. \SL, \citealt{CidFernandes_etal_2005}) to infer the physical properties of galaxies, regardless of their spectral and morphological type (e.g. \citealt{Kauffmann_etal_2003a, Panter_Heavens_Jimenez_2003, CidFernandes_etal_2005, Asari_etal_2007, Tojeiro_etal_2009}).
	
	Comparing the physical properties inferred by the population synthesis codes \Fado\ and \SL\ for four distinct galaxy samples: Main Sample (i.e. general population of galaxies), Star-Forming (EW(H$\alpha$)>3), Intensively Star-Forming (EW(H$\alpha$)>75), and Passive (EW(H$\alpha$)<0.5), shows that:
	
\begin{itemize}
\item[$\bullet$] Mass distributions for the different galaxy samples are similar between \Fado\ and \SL .

\item[$\bullet$]  Mean age and mean metallicity distributions in SFS from \Fado\ are broader than those of \SL, especially when weighted by mass. Moreover, the average light-weighted age of \SL\ is lower by $\sim$0.17 dex than \Fado\ for ISFS galaxies, whereas the light-weighted metallicity of \SL\ is $\sim$0.13 dex higher than \Fado .

\item[$\bullet$] Even though both codes show very similar mass, age and metallicity distributions for the PS population of galaxies, \Fado\ displays a narrower and higher mass-weighted age peak than \SL .

\item[$\bullet$] Cumulative light distributions of the best-fit PVs as a function of age show for ISFS that the contribution of SSPs with $t < 10^7$ yr is greater by 5.41\% in \SL\ than in \Fado , reaching 9.11\% for SSPs younger than $t < 10^9$ yr, which means that \SL\ overestimates the relative light contribution of young stellar populations in comparison with \Fado\ (or vice versa). Moreover, cumulative light distributions as a function of metallicity indicate that the light difference between \Fado\ and \SL\ for ISFS is $\sim$8.05\% and 13.51\% for $Z_\star \leq Z_\odot/200$ and $Z_\odot/50$ (i.e. the lowest metallicities in the adopted stellar library), respectively. This means that \Fado\ underestimates the metallicity when compared to \SL\ (or vice versa). Meanwhile, mass distributions as a function of age and metallicity from both codes are very similar for all samples, except that \SL\ once again fits more SSPs with ages $t<10^9$ yr than \Fado\ for both SFS and ISFS. 

\item[$\bullet$] Comparing the different relation permutations between total mass, mean age and mean metallicity shows all previous results in a more general way: (a) light-weighted age and metallicity differences between codes increases with increasing EW(H$\alpha$) and decreasing total mass, (b) mass-weighted age and metallicity differences between codes are minimal, except for the age underestimation of \SL\ in relation to \Fado\ when it comes specifically to PS galaxies, and (c) SFS age and metallicity distributions of \Fado\ are broader than those of \SL.

\end{itemize}

	These results indicate that the nebular continuum modelling approach significantly impacts the inferred stellar properties of SF galaxies, even if the negative effects of a purely stellar modelling approach are mixed with other uncertainties and degeneracies associated with population synthesis synthesis. For instance, this work found that the modelling of the nebular continuum with \Fado\ yields steeper light-weighted mass-metallicity correlation and a flatter light-weighted mass-age correlation when compared to \SL, a purely stellar population synthesis code.  Among other potential implications, this means that the stellar populations of low-mass galaxies in the local universe with relatively high specific SFRs are both more metal poor and older than previously thought. These results are particularly relevant in light of future high-resolution spectroscopic surveys at higher redshifts, such as the 4MOST-4HS (\citealt{deJong_etal_2019}) and MOONS (\citealt{Cirasuolo_etal_2011, Cirasuolo_etal_2020, Maiolino_etal_2020}), for which the fraction of intensively SF galaxies is expected to be higher.


\begin{acknowledgements}

	The authors thank the anonymous referee for valuable comments and suggestions, and colleagues Andrew Humphrey, Isreal Matute and Tom Scott (Instituto de Astrofísica e Ciências do Espaço; IA) for engaging scientific discussions. 
	
	This work was supported by Fundação para a Ciência e a Tecnologia (FCT) through the research grants UID/FIS/04434/2019, UIDB/04434/2020 and UIDP/04434/2020.

	L.S.M.C. acknowledges support by the project `Enabling Green E-science for the SKA Research Infrastructure (ENGAGE SKA)' (reference POCI-01-0145-FEDER-022217) funded by COMPETE 2020 and FCT.

	J.M.G. is supported by the DL 57/2016/CP1364/CT0003 contract and acknowledges the previous support by the fellowships CIAAUP-04/2016-BPD in the context of the FCT project UID/FIS/04434/2013 and POCI-01-0145-FEDER-007672, and SFRH/BPD/66958/2009 funded by FCT and POPH/FSE (EC).
	
	P.P. is supported by the project "Identifying the Earliest Supermassive Black Holes with ALMA (IdEaS with ALMA)" (PTDC/FIS-AST/29245/2017). 
	
	C.P. acknowledges support from DL 57/2016 (P2460) from the `Departamento de Física, Faculdade de Ciências da Universidade de Lisboa'.

	A.P.A. acknowledges support from the Fundação para a Ciência e a Tecnologia (FCT) through the work Contract No. 2020.03946.CEECIND, and through the FCT project EXPL/FIS-AST/1085/2021.

	J.A. acknowledges financial support from the Science and Technology Foundation (FCT, Portugal) through research grants UIDB/04434/2020 and UIDP/04434/2020.

	P.L. is supported by the DL 57/2016/CP1364/CT0010 contract.


	Funding for the SDSS and SDSS-II has been provided by the Sloan Foundation, the Participating Institutions, the National Science Foundation, the U.S. Department of Energy, the National Aeronautics and Space Administration, the Japanese Monbukagakusho, the Max Planck Society, and the Higher Education Funding Council for England.
	The SDSS Web Site is \url{http://www.sdss.org/}. The SDSS is managed by the Astrophysical Research Consortium for the Participating Institutions. The Participating Institutions are the American Museum of Natural History, Astrophysical Institute Potsdam, University of Basel, University of Cambridge, Case Western Reserve University, University of Chicago, Drexel University, Fermilab, the Institute for Advanced Study, the Japan Participation Group, Johns Hopkins University, the Joint Institute for Nuclear Astrophysics, the Kavli Institute for Particle Astrophysics and Cosmology, the Korean Scientist Group, the Chinese Academy of Sciences (LAMOST), Los Alamos National Laboratory, the Max-Planck-Institute for Astronomy (MPIA), the Max-Planck-Institute for Astrophysics (MPA), New Mexico State University, Ohio State University, University of Pittsburgh, University of Portsmouth, Princeton University, the United States Naval Observatory, and the University of Washington. 

\end{acknowledgements}



\normalsize

\clearpage

\begin{appendix}

\section{Supplements}\label{Appendix_-_Supplements}

\begin{figure*}
\begin{center}	
\includegraphics[width=\textwidth]{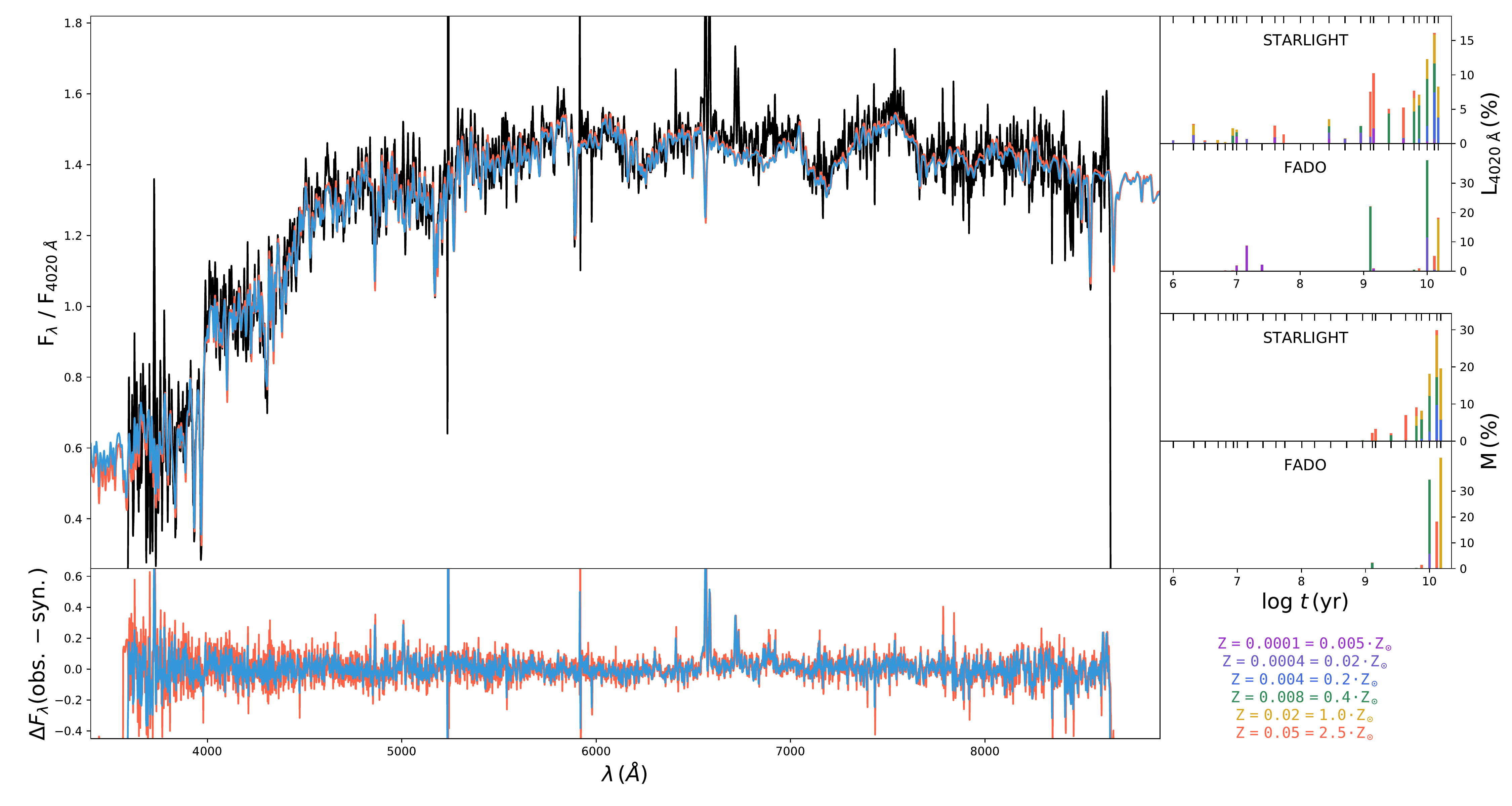}
\caption{Spectral modelling results from \SL\ and \Fado\ for the SDSS spectrum `51630-0266-004' in the Main Sample. The format and other legend details are identical to those in Fig. \ref{Fig_-_text_fit_example}.}
\label{Fig_-_MS_-_fit_example_1}
\end{center}
\end{figure*}

\begin{figure*}
\begin{center}	
\includegraphics[width=\textwidth]{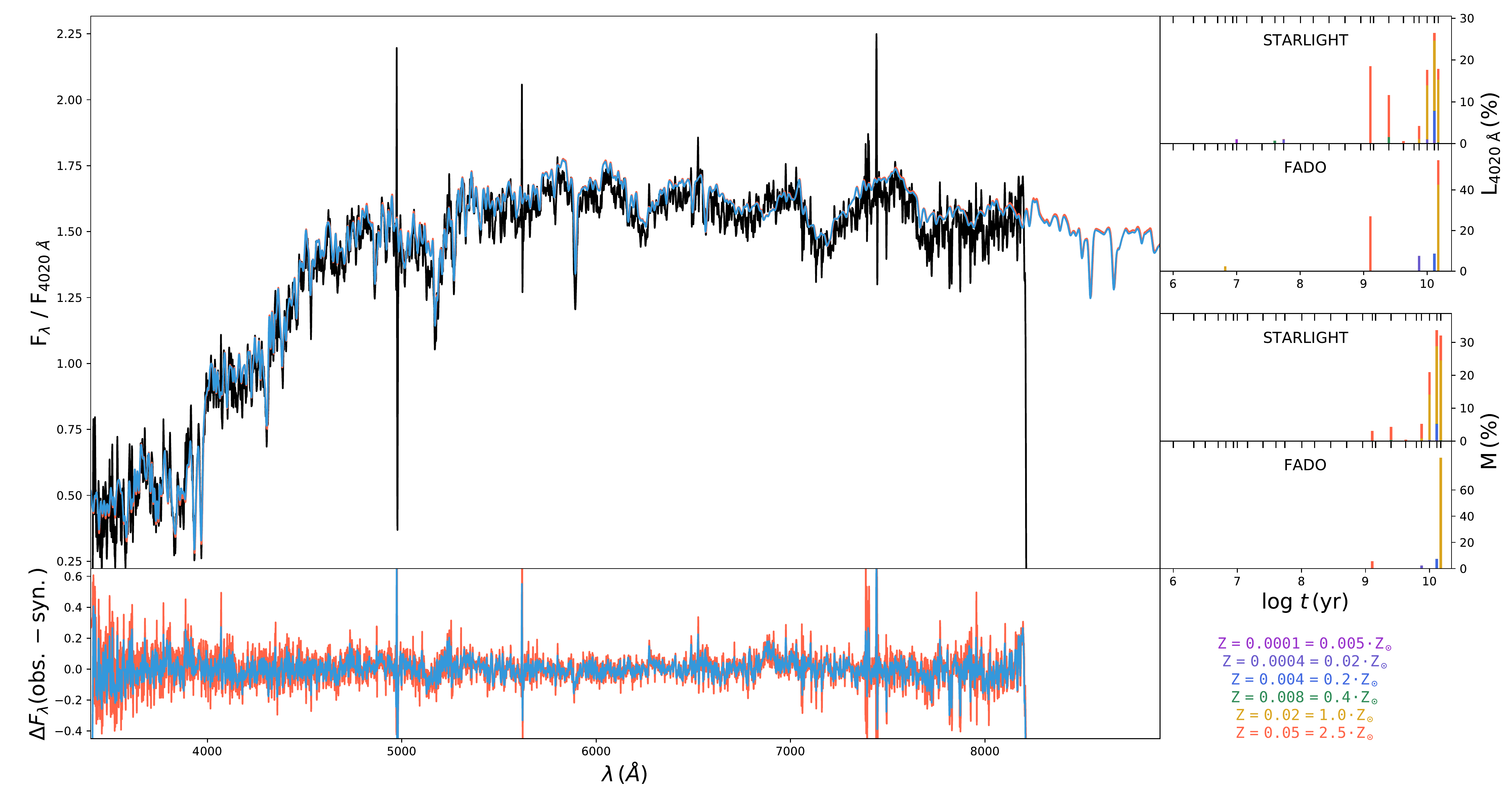}
\caption{Spectral modelling results from \SL\ and \Fado\ for the SDSS spectrum `51630-0266-011' in the Main Sample. The format and other legend details are identical to those in Fig. \ref{Fig_-_text_fit_example}.}
\label{Fig_-_MS_-_fit_example_2}
\end{center}
\end{figure*}

\begin{figure*}
\begin{center}	
\includegraphics[width=\textwidth]{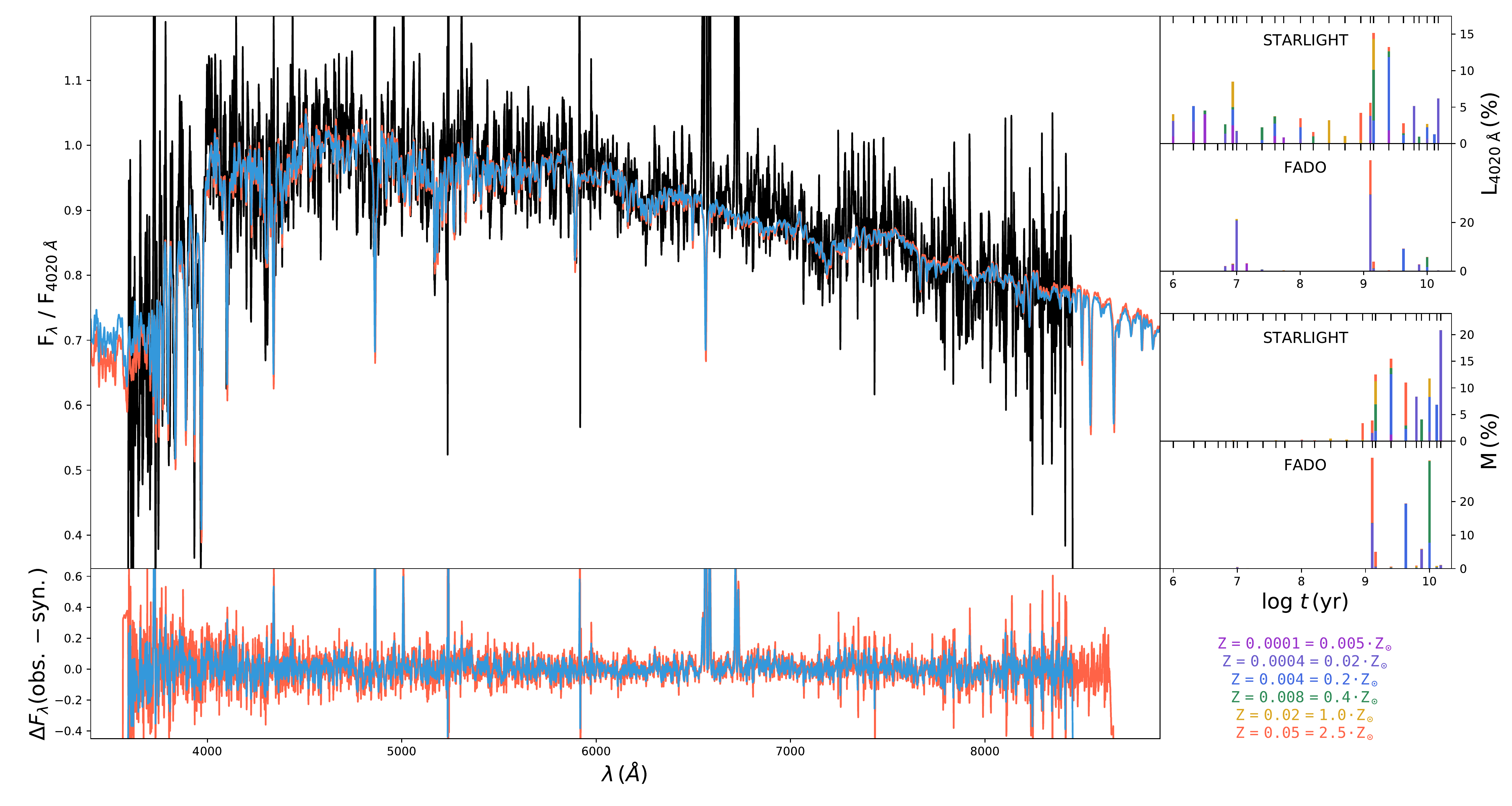}
\caption{Spectral modelling results from \SL\ and \Fado\ for the SDSS spectrum `51630-0266-014' in the Star-Forming Sample. The format and other legend details are identical to those in Fig. \ref{Fig_-_text_fit_example}.}
\label{Fig_-_SFS_-_fit_example_1}
\end{center}
\end{figure*}

\begin{figure*}
\begin{center}	
\includegraphics[width=\textwidth]{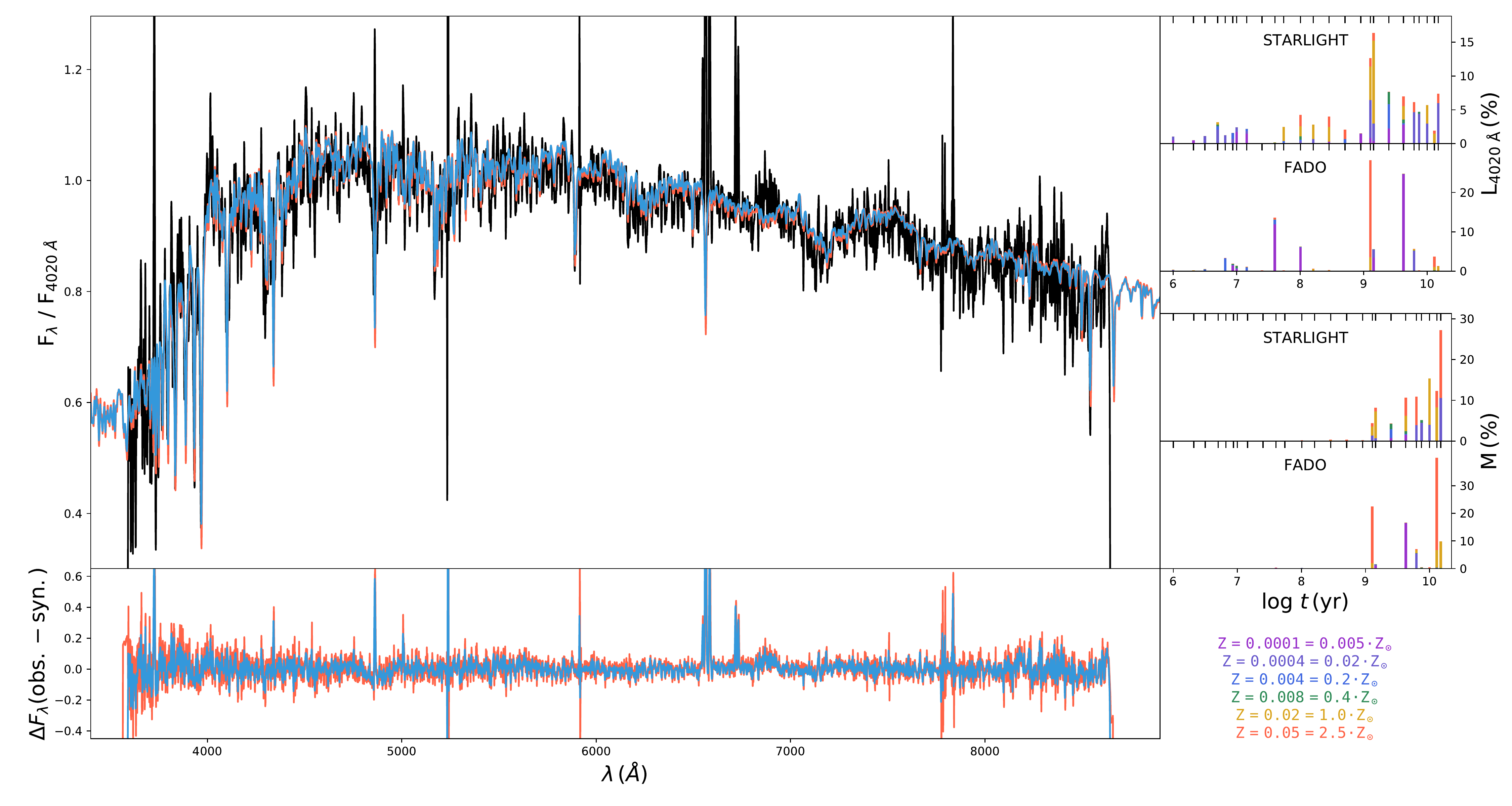}
\caption{Spectral modelling results from \SL\ and \Fado\ for the SDSS spectrum `51630-0266-017' in the Star-Forming Sample. The format and other legend details are identical to those in Fig. \ref{Fig_-_text_fit_example}.}
\label{Fig_-_SFS_-_fit_example_2}
\end{center}
\end{figure*}

\begin{figure*}
\begin{center}	
\includegraphics[width=\textwidth]{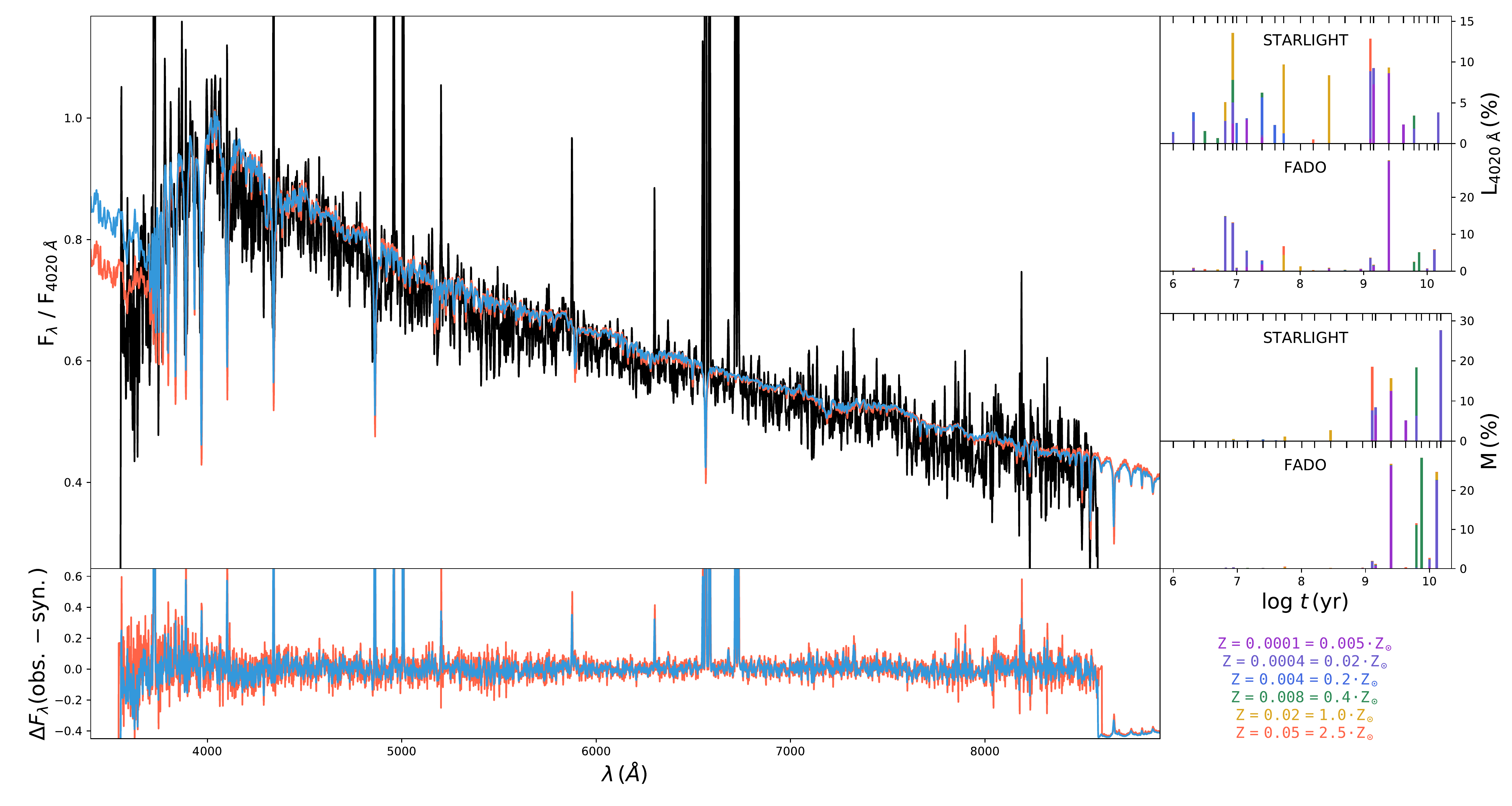}
\caption{Spectral modelling results from \SL\ and \Fado\ for the SDSS spectrum `51957-0273-573' in the Intensively Star-Forming Sample. The format and other legend details are identical to those in Fig. \ref{Fig_-_text_fit_example}.}
\label{Fig_-_ISFS_-_fit_example_1}
\end{center}
\end{figure*}

\begin{figure*}
\begin{center}	
\includegraphics[width=\textwidth]{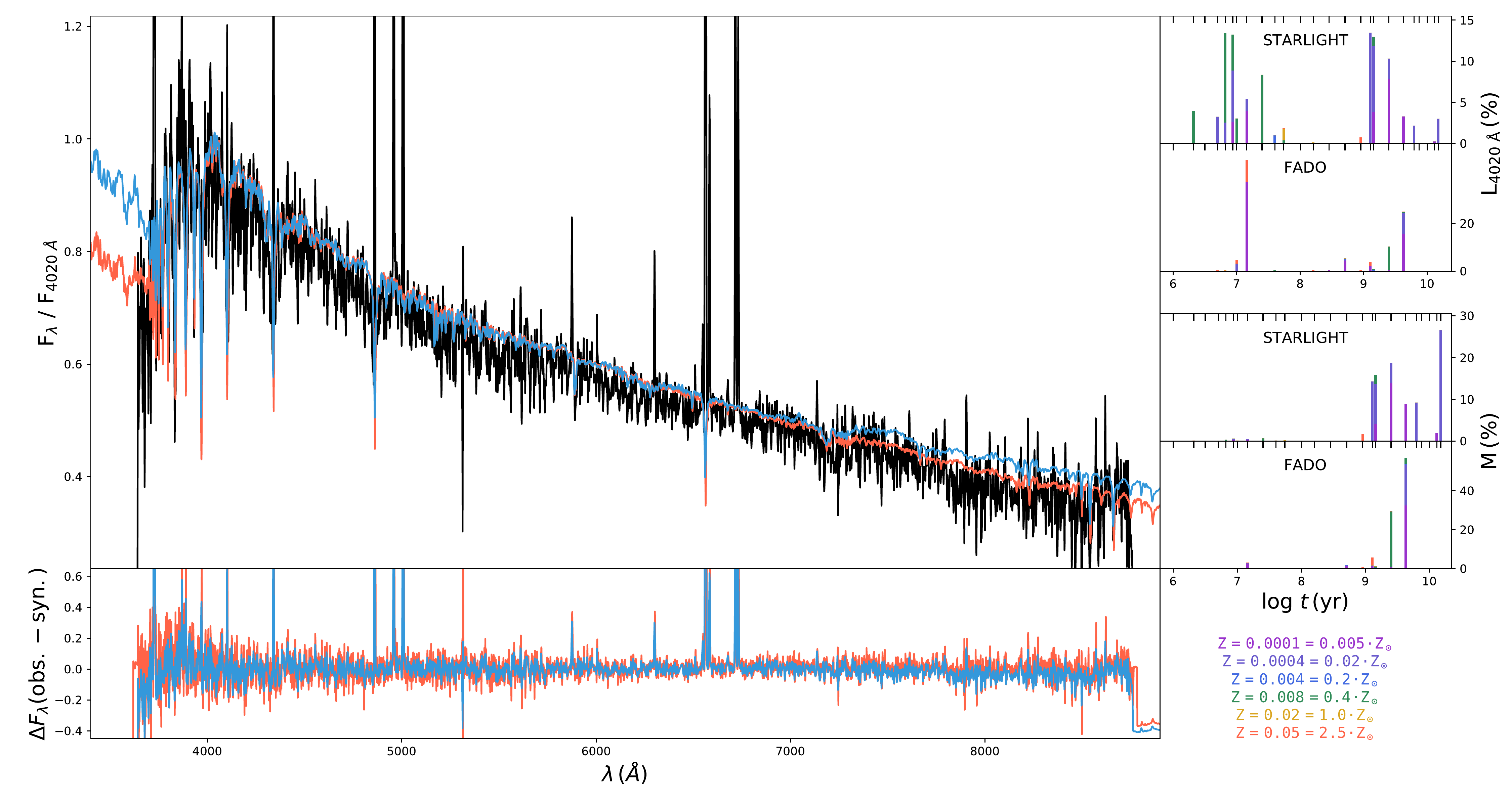}
\caption{Spectral modelling results from \SL\ and \Fado\ for the SDSS spectrum `51957-0273-625' in the Intensively Star-Forming Sample. The format and other legend details are identical to those in Fig. \ref{Fig_-_text_fit_example}.}
\label{Fig_-_ISFS_-_fit_example_2}
\end{center}
\end{figure*}

\begin{figure*}
\begin{center}	
\includegraphics[width=\textwidth]{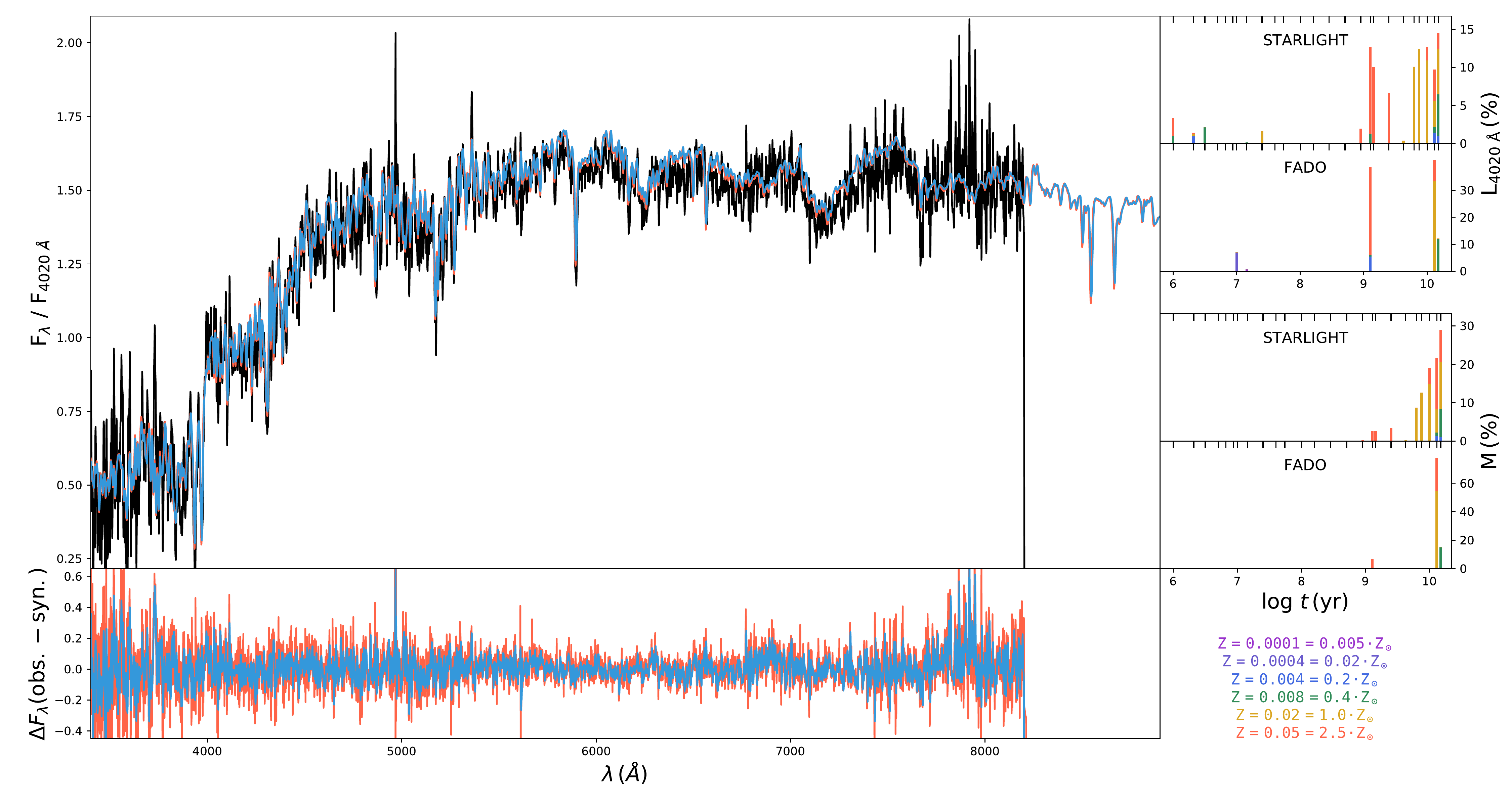}
\caption{Spectral modelling results from \SL\ and \Fado\ for the SDSS spectrum `51630-0266-491' in the Passive Sample. The format and other legend details are identical to those in Fig. \ref{Fig_-_text_fit_example}.}
\label{Fig_-_PS_-_fit_example_1}
\end{center}
\end{figure*}

\begin{figure*}
\begin{center}	
\includegraphics[width=\textwidth]{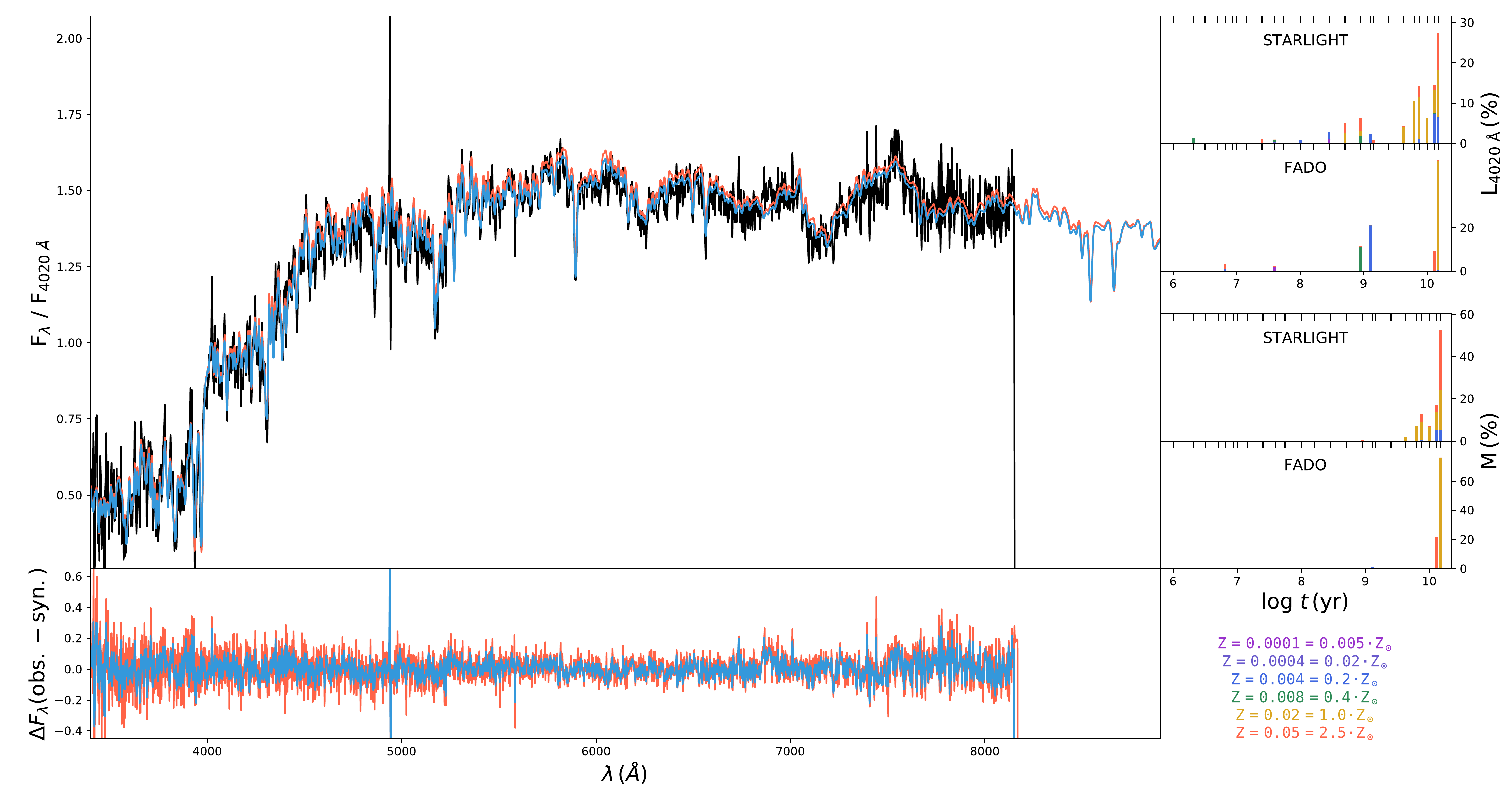}
\caption{Spectral modelling results from \SL\ and \Fado\ for the SDSS spectrum `51630-0266-493' in the Passive Sample. The format and other legend details are identical to those in Fig. \ref{Fig_-_text_fit_example}.}
\label{Fig_-_PS_-_fit_example_2}
\end{center}
\end{figure*}

	Figures \ref{Fig_-_MS_-_fit_example_1}--\ref{Fig_-_PS_-_fit_example_2} show the results of the \Fado\ and \SL\ spectral fits for two randomly selected spectra from each galaxy population defined in Subsection \ref{SubSection_-_Sample_Selection}: Main Sample, Star-Forming Sample, Intensively Star-Forming Sample and Passive Sample, respectively. The light-weighted SFH results for the SFS (Figures \ref{Fig_-_SFS_-_fit_example_1} \& \ref{Fig_-_SFS_-_fit_example_2}) and ISFS spectra (Figures \ref{Fig_-_ISFS_-_fit_example_1} \& \ref{Fig_-_ISFS_-_fit_example_2}) show in particular the differences between \Fado\ and \SL, with the relative contribution of SSPs younger than 100 Myr in \SL\ being much more prominent than those in the \Fado\ results.

\begin{figure*}[!t] 
\begin{center}	
\includegraphics[width=0.9\textwidth]{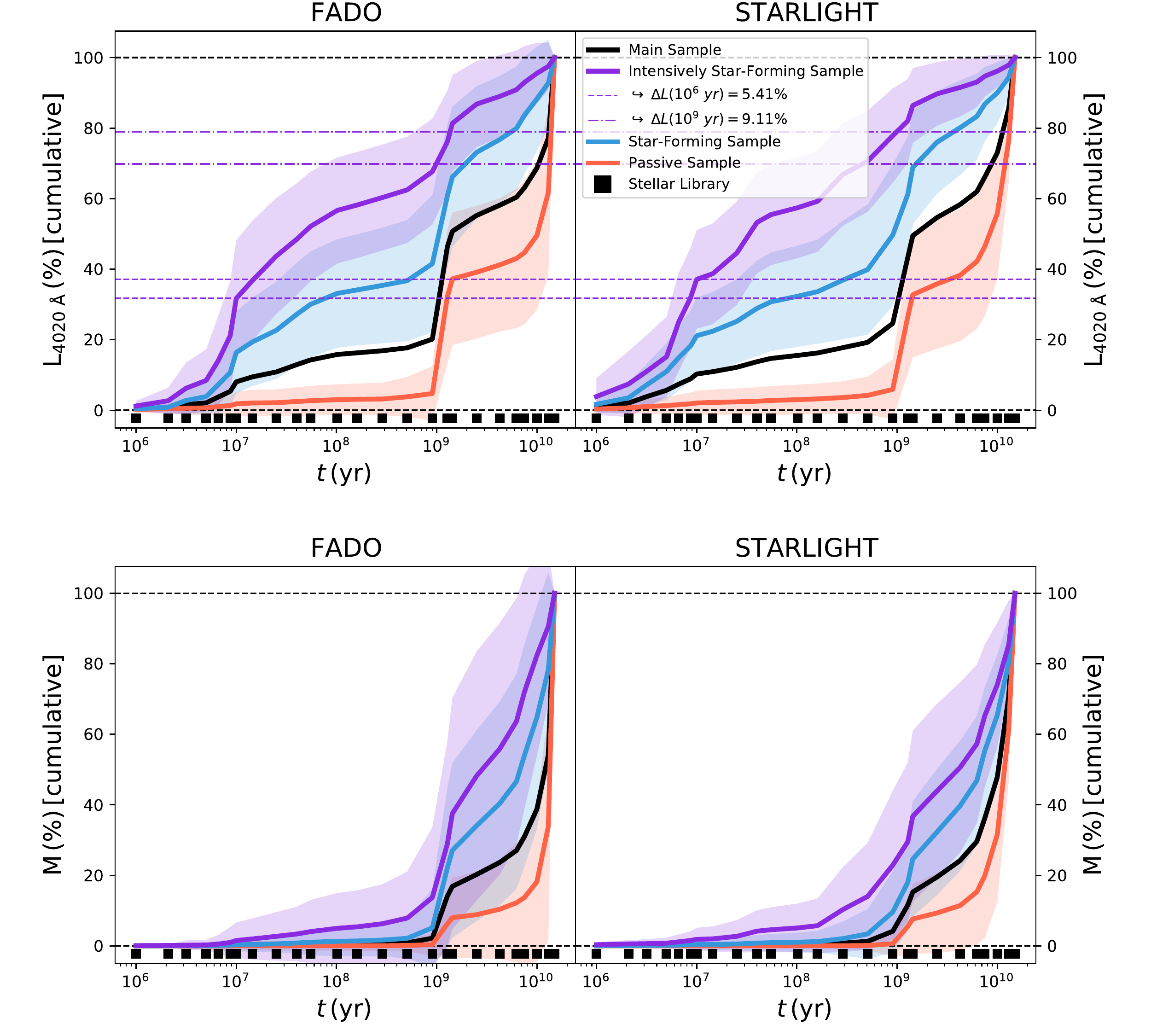}
\caption{ Cumulative light (top panels) and mass (bottom panels) relative contributions of the stellar library elements as a function of their age. Other legend details are identical to those in Fig. \ref{Fig_-_SFH_vs_age}.}
\label{Fig_-_SFH_vs_age_-_cumulative}
\end{center}
\end{figure*}

\begin{figure*}[!t] 
\begin{center}	
\includegraphics[width=0.9\textwidth]{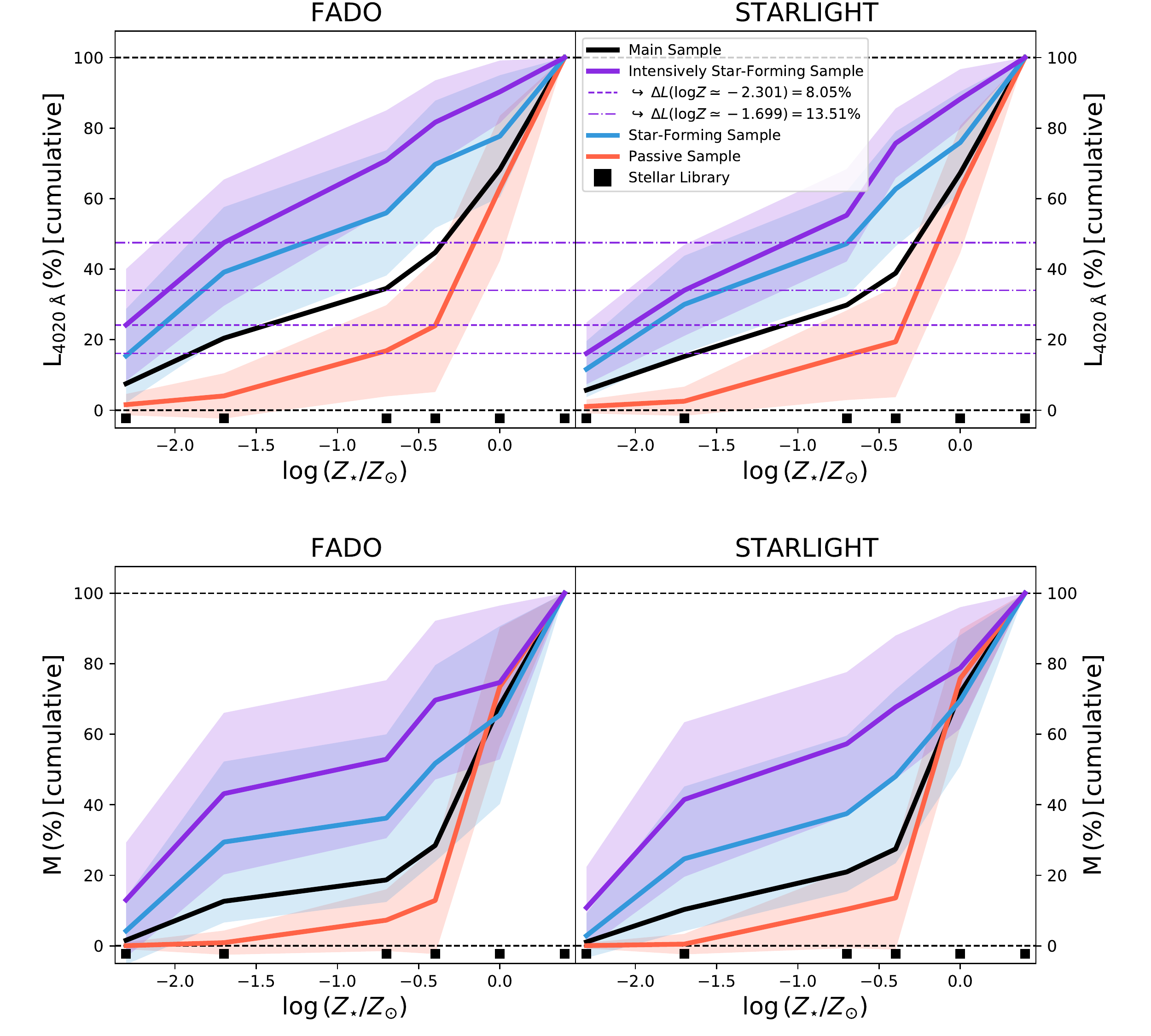}
\caption{ Cumulative light (top panels) and mass (bottom panels) relative contributions of the stellar library elements as a function of their  metallicity. Other legend details are identical to those in Fig. \ref{Fig_-_SFH_vs_Z}.}
\label{Fig_-_SFH_vs_Z_-_cumulative}
\end{center}
\end{figure*}

	Figures \ref{Fig_-_SFH_vs_age_-_cumulative} and \ref{Fig_-_SFH_vs_Z_-_cumulative} show the cumulative versions of Figures \ref{Fig_-_SFH_vs_age} and \ref{Fig_-_SFH_vs_Z} with the light and mass contributions being summed with increasing age and metallicity, respectively. The fact that the assembly histories for the PS are practical identical between codes is an important sign that both codes estimate similar SFHs for galaxies with negligible to none nebular contribution to the continuum (\EWha <0.5 \AA).

\begin{figure*}[!t] 
\begin{center}	
\includegraphics[width=\textwidth]{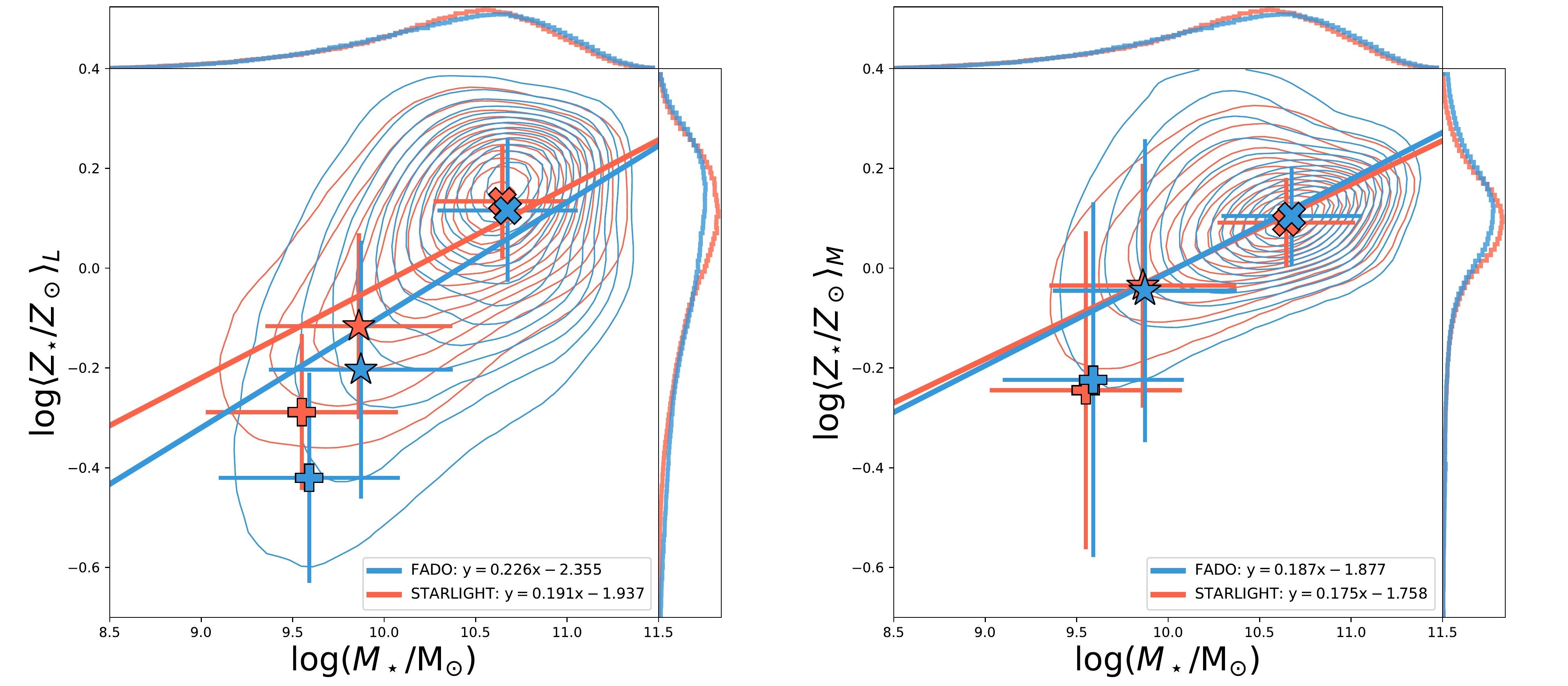}
\caption{ Mean stellar metallicity and $\log \langle Z_\star \rangle$ as a function of total stellar mass $M_\star$for the Main Sample. Blue and red lines and points represent \Fado\ and \SL\ results, respectively. The `\ding{58}', `$\bigstar$' and `\ding{54}' symbols represent the average values for the ISFS, SFS and PS populations, respectively, with standard deviation errorbars. Linear regression results for the MS population from each code are present in the legend of each plot.}
\label{Fig_-_MS_-_M_vs_Z}
\end{center}
\end{figure*}

\begin{figure*}[!t] 
\begin{center}	
\includegraphics[width=\textwidth]{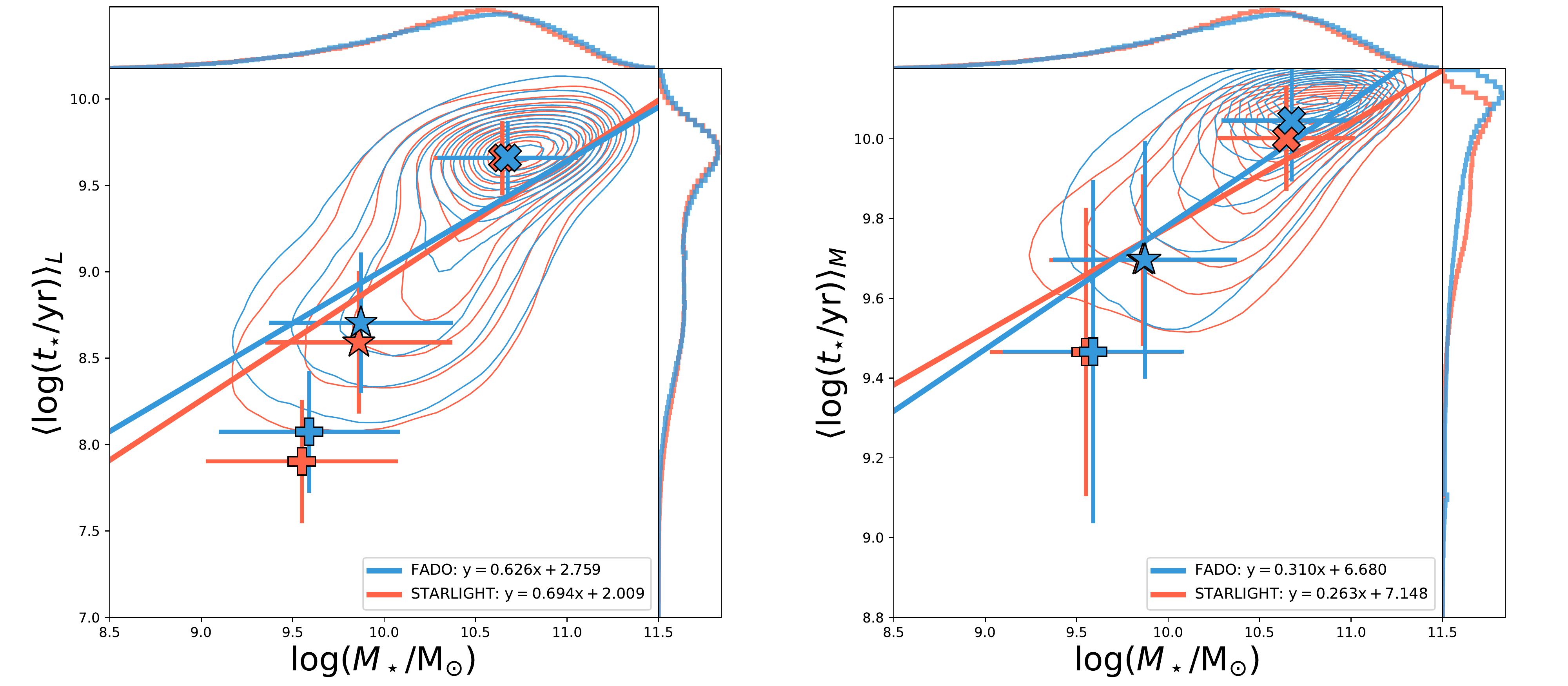}
\caption{ Mean stellar age $\langle \log t_\star \rangle$ as a function of total stellar mass $M_\star$ for the Main Sample. Other legend details are identical to those in Fig. \ref{Fig_-_MS_-_M_vs_Z}.}
\label{Fig_-_MS_-_M_vs_t}
\end{center}
\end{figure*}

\begin{figure*}[!t] 
\begin{center}	
\includegraphics[width=\textwidth]{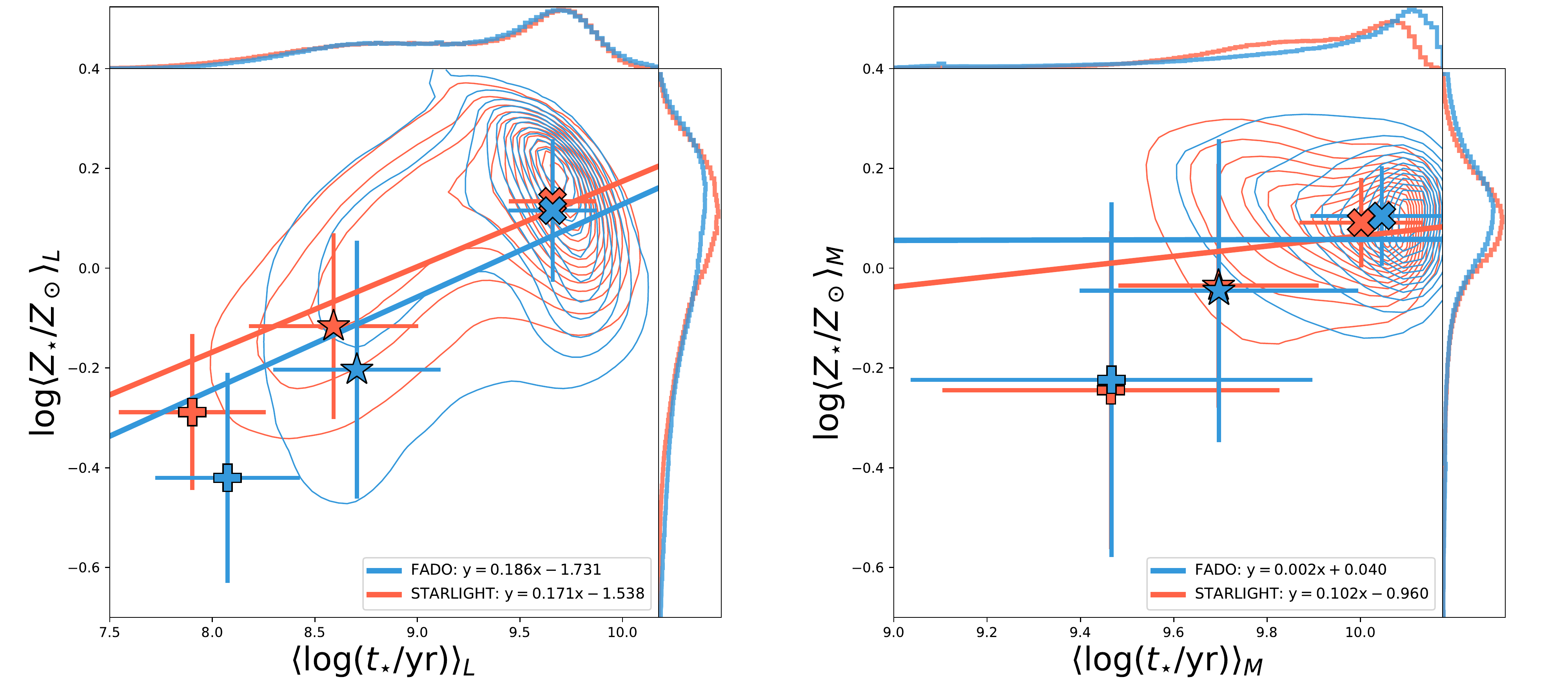}
\caption{ Mean stellar metallicity and $\log \langle Z_\star \rangle$ as a function of mean stellar age $\langle \log t_\star \rangle$ for the Main Sample. Other legend details are identical to those in Fig. \ref{Fig_-_MS_-_M_vs_Z}.}
\label{Fig_-_MS_-_t_vs_Z}
\end{center}
\end{figure*}

	Figures \ref{Fig_-_MS_-_M_vs_Z}, \ref{Fig_-_MS_-_M_vs_t} and \ref{Fig_-_MS_-_t_vs_Z} shows the relations between the total stellar mass, mean stellar age and mean stellar metallicity for the MS of galaxies for both codes. The contour and histograms show that the distributions for the three stellar properties in both codes are similar, with the only interesting difference being that \Fado\ displays a slightly larger light-weighted mean metallicity distribution than \SL.

\begin{figure*}[!t] 
\begin{center}	
\includegraphics[width=0.9\textwidth]{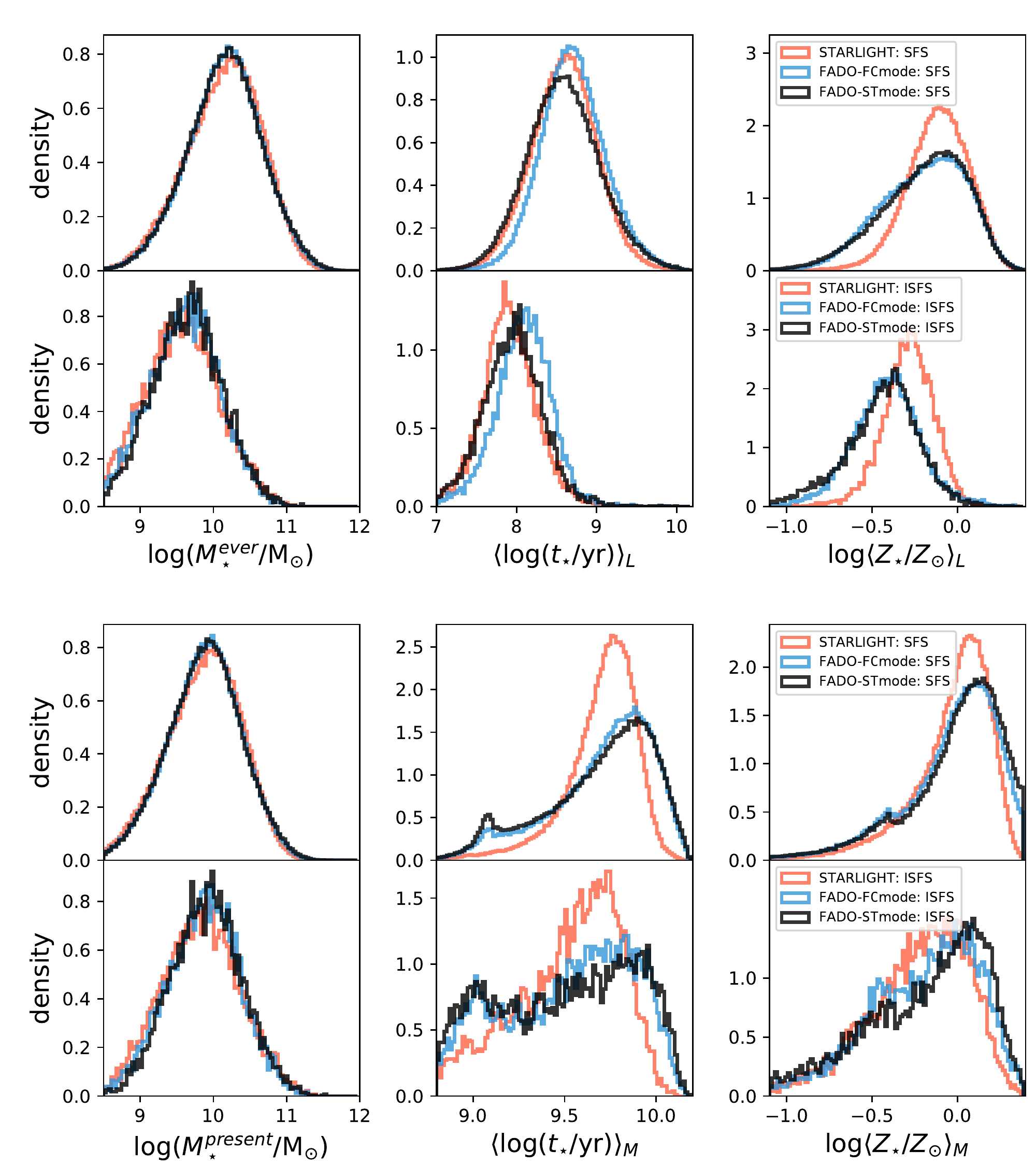}
\caption{Density distributions of the total stellar mass $M_\star$ in solar units $M_{\odot}$ (left-hand side panels), mean stellar age $\langle \log t_\star \rangle$ in years (centred panels) and mean stellar metallicity $\log \langle Z_\star \rangle$ in solar units (right-hand side panels). Red, blue and black lines represent results from \SL\ and \Fado\ in `full-consistency mode' ($\mathtt{FC}$mode) and `purely stellar mode' ($\mathtt{ST}$mode), respectively. Moreover, top and bottom row panels in each section represent results for the Star-Forming (SFS) and Intensively Star-Forming Samples (ISFS), respectively.}
\label{Fig_-_MtZ_Histograms_-_PurelyStellar}
\end{center}
\end{figure*}
\begin{figure*}[!t] 
\begin{center}	
\includegraphics[width=0.9\textwidth]{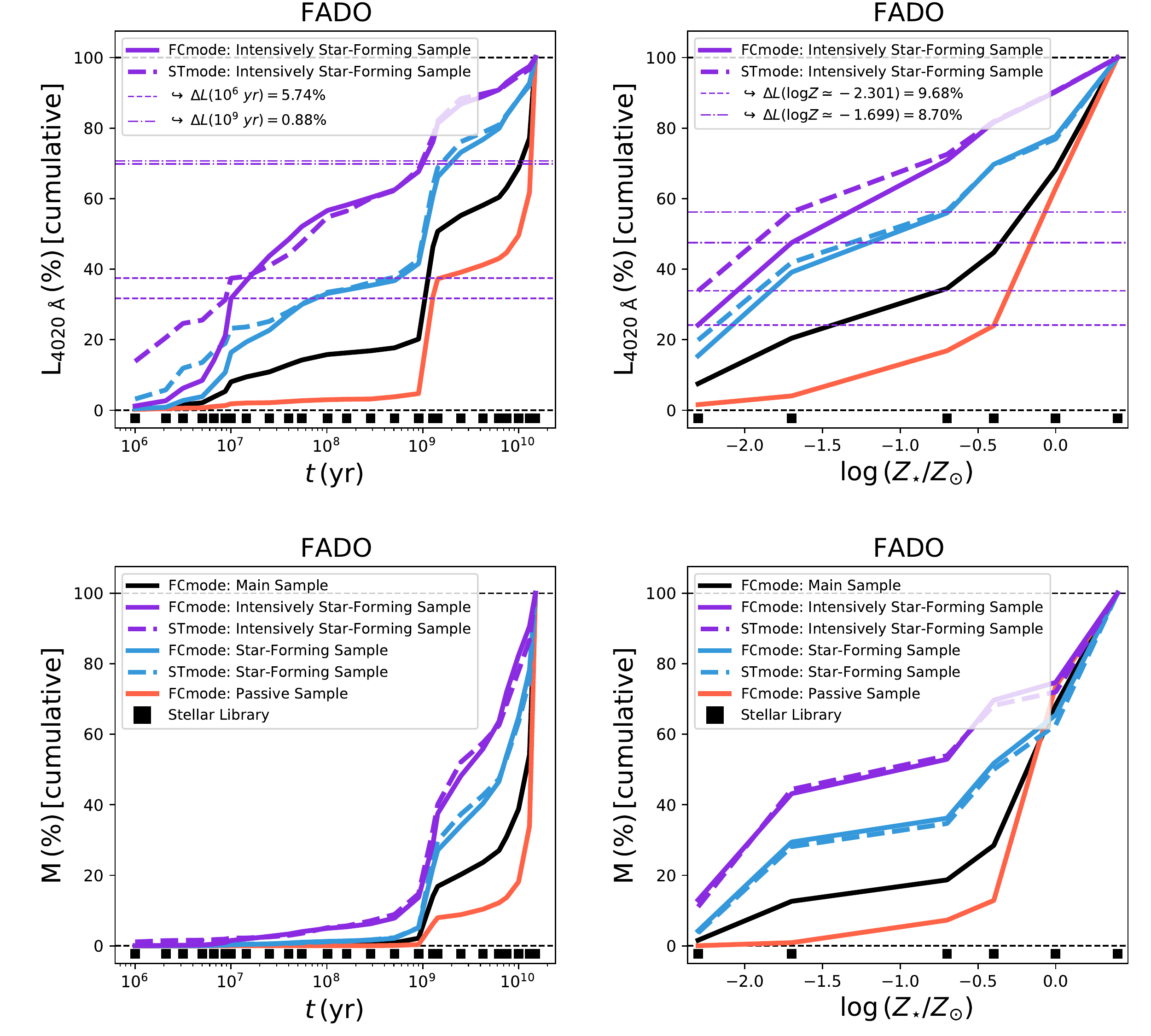}
\caption{ Cumulative light (top panels) and mass (bottom panels) relative contributions of the stellar library elements as a function of their age (left-hand side panels) and metallicity (right-hand side panels). Full and dashed lines represent \Fado\ results in `full-consistency mode' ($\mathtt{FC}$mode) and `purely stellar mode' ($\mathtt{ST}$mode), respectively. Black squares on the left and right-hand side panels represent the age and metallicity coverages of the adopted stellar library, respectively. Other legend details are identical to those in Fig. \ref{Fig_-_SFH_vs_age}.}
\label{Fig_-_SFH_vs_age_&_Z_-_PurelyStellar_-_cumulative}
\end{center}
\end{figure*}
\end{appendix}

\end{document}